\documentclass[nonblindrev,opre]{informs3_ejor} % current default for manuscript submission

\OneAndAHalfSpacedXI % current default line spacing
\usepackage{xcolor}
\usepackage[]{hyperref}% -> Set to draft mode before submission
\RequirePackage{fix-cm} 
\usepackage[]{algorithm2e}
\usepackage{amsmath}
\usepackage{xspace}
	\newcommand\ie{i.\,e.\xspace}
	\newcommand\eg{e.\,g.\xspace}

	\usepackage{siunitx}
    \DeclareSIUnit\eur{\officialeuro}
    \DeclareSIUnit\M{M}
    \DeclareSIUnit\k{k}
	
	\usepackage[%
  sort&compress
]{cleveref}
  \crefname{chapter}{section}{sections}
	\Crefname{chapter}{Section}{Sections}

\usepackage{etoolbox}
\robustify\bfseries

\usepackage{enumitem}

\hypersetup{
	citebordercolor=white,
	linkbordercolor=white,
  urlbordercolor=blue}

\usepackage{isomath}
\usepackage{bm}

\usepackage{colortbl}
\usepackage{tabularx}
\usepackage{booktabs}
\usepackage{collcell}
\usepackage{subfig}

\usepackage{float}
\usepackage{rotating}

\usepackage{array}
\newcolumntype{L}[1]{>{\raggedright\let\newline\\\arraybackslash\hspace{0pt}}p{#1}}
\newcolumntype{C}[1]{>{\centering\let\newline\\\arraybackslash\hspace{0pt}}p{#1}}
\newcolumntype{R}[1]{>{\raggedleft\let\newline\\\arraybackslash\hspace{0pt}}p{#1}}

\usepackage{csquotes}
\usepackage{amsmath}
\usepackage{amssymb}
\usepackage{pifont}

\usepackage[
    colorinlistoftodos,
    textsize=footnotesize,
        ]{todonotes}

\makeatletter
    \renewcommand{\fps@figure}{H}         % default {tbp}
    \renewcommand{\fps@table}{H}         % default {tbp}
\makeatother 

\usepackage{float}
\usepackage{rotating}

\makeatletter
\newcolumntype{B}[3]{>{\boldmath\DC@{#1}{#2}{#3}}c<{\DC@end}}
\makeatother

\usepackage{natbib}
 \bibpunct[, ]{(}{)}{,}{a}{}{,}%
 \def\bibfont{\small}%
\TheoremsNumberedThrough     % Preferred (Theorem 1, Lemma 1, Theorem 2)
\EquationsNumberedThrough    % Default: (1), (2), ...
\newcommand{\model}{\mbox{VDC-HMMX}\xspace}

\begin{document}
%%%%%%%%%%%%%%%%

\RUNTITLE{Data-driven dynamic treatment planning for chronic diseases}

\TITLE{Data-driven dynamic treatment planning for chronic diseases}

\ARTICLEAUTHORS{%
\AUTHOR{Christof Naumzik}
\AFF{ETH Zurich, Weinbergstr. 56/58, 8092 Zurich, Switzerland, \EMAIL{cnaumzik@ethz.ch}, \URL{}}
\AUTHOR{Stefan Feuerriegel (corresponding author)}
\AFF{LMU Munich, Geschwister-Scholl-Platz 1, 80539 Munich, Germany, \EMAIL{feuerriegel@lmu.de}, \URL{}}
\AUTHOR{Anne Molgaard Nielsen}
\AFF{Department of Sports Science and Clinical Biomechanics, University of Southern Denmark, Denmark, \EMAIL{amnielsen@health.sdu.dk}, \URL{}}
} % end of the block

\ABSTRACT{%
%\SingleSpacedXI\small
In order to deliver effective care, health management must consider the distinctive trajectories of chronic diseases. These diseases recurrently undergo acute, unstable, and stable phases, each of which requires a different treatment regimen. However, the correct identification of trajectory phases, and thus treatment regimens, is challenging. In this paper, we propose a data-driven, dynamic approach for identifying trajectory phases of chronic diseases and thus suggesting treatment regimens. Specifically, we develop a novel \emph{variable-duration copula hidden Markov model} (\model). In our \model, the trajectory is modeled as a series of latent states with acute, stable, and unstable phases, which are eventually recovered. We demonstrate the effectiveness of our \model model on the basis of a longitudinal study with 928 patients suffering from low back pain. A myopic classifier identifies correct treatment regimens with a balanced accuracy of slightly above 70\%. In comparison, our \model model is correct with a balanced accuracy of 83.65\%. This thus highlights the value of longitudinal monitoring for chronic disease management.  
}

% chronic diseases; 

\HISTORY{}

\maketitle

\sloppy
\raggedbottom

%%%%%%%%%%%%%%%%%%%%%%%%%%%%%%%%%%%%%%%%%%%%%%%%%%%%%%%%%%%%%%%%%%%%%%
\vspace{-1cm}
\section{Introduction}

Chronic diseases, such as depression, arthritis, cancer, diabetes, and non-specific low back pain, represent a considerable burden upon individuals and society. In the US alone, chronic diseases affect the quality of life of about 40 million individuals, nearly one third of the adult population \citep{Blackwell.2014}. Owing to this prevalence, there is an urgent need for better chronic disease management. In particular, over- and under-treatment are critical, as these lead to suboptimal care \citep{over_under_treatment}.

Chronic disease management must be carefully adapted to the distinctive characteristics of chronic diseases. Chronic diseases are health conditions that are long-lasting or persistent \citep{Bernell.2016}. Furthermore, chronic diseases undergo different -- and potentially recurrent -- trajectory phases comprising acute, unstable, and stable periods \citep{Corbin.1991,Corbin.1998}. In each of these trajectory phases, a chronic disease is characterized by different disease dynamics. It also demands different forms of care and thus requires tailored treatment regimens. If the trajectory phase changes, the treatment regimen must be updated accordingly. This has been formalized in the so-called chronic illness trajectory framework \citep{Corbin.1988,Corbin.1991}.  

\begin{figure}[htbp]%[H]
\label{fig:problem_statement}
	\centering
	\includegraphics[width=0.8\textwidth]{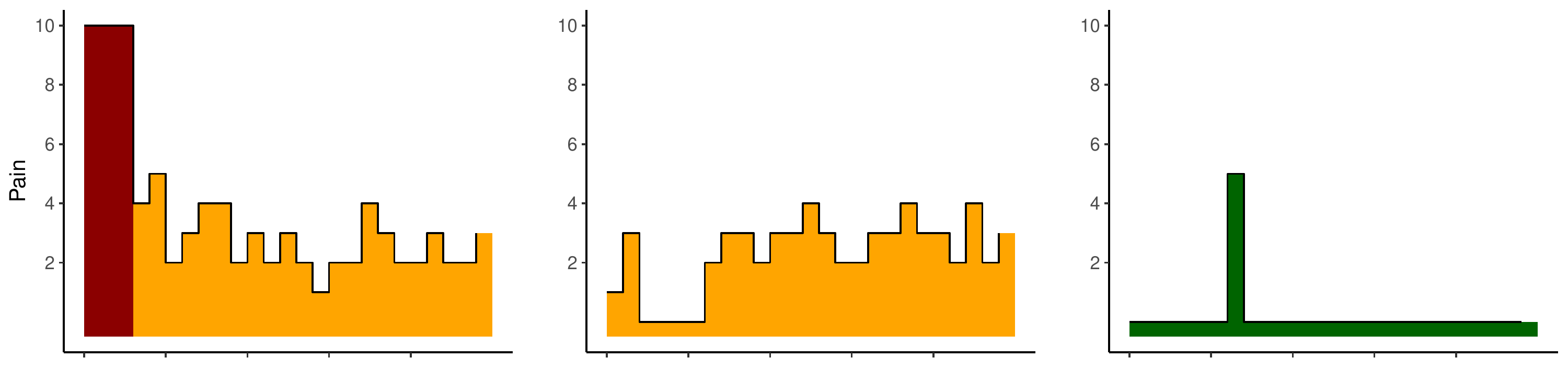}
\baselineskip10pt plus 0.4pt minus 0.3pt
\begin{tabular}{p{16cm}}
\scriptsize\emph{Notes}: The plot shows the disease progression of three sample patients. The trajectory phases were annotated by medical experts and are shown in color, namely an acute trajectory phase in red, unstable in yellow, and stable in green. \emph{Left:} The patient exhibits a quick recovery from a temporary relapse. \emph{Center:} The patient has a temporary absence of symptoms, and yet a treatment regimen for an unstable trajectory phase would be preferable overall. \emph{Right:} The patient experiences a sudden pain episode. By focusing on symptoms, a health professional might be inclined to recommend an acute treatment regimen (\eg, strong medicine with potential side-effects). Yet the symptoms would revert quickly and, hence, a stable treatment regimen (\eg, pain self-management) appears desirable. In all three examples, the trajectory phase is a better indicator of the true progression of the disease and, therefore, the guidelines from the trajectory framework \citep{Corbin.1991} advise basing treatment decisions on the trajectory phase (rather than on symptoms). Note that the above examples draw upon the complete course and thus benefit from post-hoc knowledge, whereas, in practice, the course would only be known partially. Hence, identifying a suitable treatment regimen in such a dynamic setting is challenging. 

\caption{Example Health Trajectories for Low Back Pain.}
\end{tabular}
\end{figure}

Applying the above trajectory framework in practice is challenging. Despite this guideline being regarded as best practice \citep{Larsen.2017}, healthcare practitioners often adapt care to symptoms rather than the underlying trajectory \citep{Snyderman.2012}. The primary reason is that the identification of trajectory phases is non-trivial \citep{Burton.2000}. For one thing, trajectory phases and symptoms are only stochastically related. For instance, there might be a temporary change in symptoms, while the underlying trajectory phase remains unaffected (or vice versa). Hence, drawing inferences from symptoms rather than trajectory phases is sub-optimal \citep{Corbin.1988,Corbin.1991}. This is shown on the basis of example health trajectories in \Cref{fig:problem_statement}. Moreover, the course of symptoms is highly variable. Some variation in symptoms is thus stochastic in nature and should quickly revert to its original condition. Hence, it is often difficult to identify the onset of a new trajectory phase and thus when a new treatment regimen should be applied. 

This work develops a data-driven approach to dynamic treatment planning for chronic diseases. For this purpose, we formed an interdisciplinary team with health researchers and designed our data-driven approach according to the trajectory framework: it dynamically determines the (latent) trajectory phase and, based on it, suggests an acute, unstable, or stable treatment regimen. Formally, we develop a novel \textbf{variable-duration copula hidden Markov model~(\model)}. In our \model, the health trajectory forms a a sequence of latent states (acute, stable, unstable), for which symptoms represent noisy realizations. Our \model then recovers the latent trajectory phases, which then prescribe the treatment regimens. 

Thereby, our model extends the na{\"i}ve HMM \citep{MacDonald.1997,Rabiner.1989} in three ways: (1)~a variable duration component, (2)~a copula structure for multivariate emissions, and (3)~heterogeneity in the transition. The latter includes risk variables that describe the between-patient heterogeneity. Previously, several works have used \emph{either} variable-duration HMMs \citep[\eg,][]{Barbu.2008,Chiappa.2014,Kundu.1998,Limnios.2008,Murphy.2002,MKSC,Yu.2010} \emph{or} copula HMMs \citep[\eg,][]{Brunel.2005,Brunel.2010,CopulaHMM1,Martino.2018,CopulaHMM2}. Both variable-duration HMM and copula HMM are later part of our baselines and outperformed by our proposed \model. However, to the best of our knowledge, no work has hitherto developed a HMM accommodating (1)--(3), thus making ours the first variable-duration copula HMM. 

% Even others have used copulas but outside of HMMs \citep[\eg,][]{Copula1,Copula2}. 

The effectiveness of our HMM is demonstrated based on a longitudinal study of almost \num[group-minimum-digits=3]{1000} patients suffering from recurrent low back pain. This disease is responsible for the greatest number of years spent with disability globally \citep{GBD.2017}. Based on a dynamic evaluation framework, we obtain the following finding. A myopic classifier (reflecting current practice) identifies the correct treatment regimen with a balanced accuracy of slightly above \SI{70}{\percent}, whereas the proposed HMM achieves a balanced accuracy of \SI{83.65}{\percent}. Our proposed HMM further outperforms a na{\"i}ve HMM as well as a sequential neural network (long short-term memory). 

Our work has direct implications for health management. First, health professionals should be careful if treatment decisions are based purely on symptoms: symptoms merely represent noisy realizations of the underlying health trajectory and thus prompt ineffective treatment regimens. Rather, the decision-making of health professionals should be aligned with the underlying trajectory phase. This latter approach provides a means by which over- and under-treatment could be alleviated. Second, health management could benefit from further personalization with regard to individual health trajectories. Third, our work encourages health professionals in chronic disease management to more widely utilize longitudinal health monitoring. 

This paper is organized as follows. \Cref{sec:background} reviews the trajectory framework as state-of-the-art practice in chronic disease management. \Cref{sec:method} develops our novel variable-duration copula HMM. \Cref{sec:study_setting} presents our longitudinal study of low back pain. We then first select our preferred HMM specification (\Cref{sec:model_selection}) and then benchmark it against baselines to demonstrate the effectiveness of our HMM for dynamically identifying treatment regimens (\Cref{sec:results}). Finally, \Cref{sec:discussion} discusses implications for personalized chronic disease management. \Cref{sec:conclusion} concludes. 

\section{Background}
\label{sec:background}

\subsection{Managing the Trajectory of Chronic Diseases}

Managing chronic diseases aims at stabilizing the underlying course \citep{Henly.2017}. To this end, chronic disease management \textquote{does not necessarily mean altering the direction of the course} \citep{Corbin.1991}, as this is usually not possible \citep{Larsen.2017}. Rather, chronic disease management aims at maintaining the current trajectory phase, as detailed in the following. 

In chronic care, health management must adapt to the protracted, variable, and often recurrent course of chronic diseases \citep{Bernell.2016}. This has been formalized in the so-called \emph{trajectory framework} \citep{Corbin.1991}. Combining 30 years of health management research, this comprehensive framework facilitates the task of managing the course of chronic diseases. The framework was later extended to incorporate additional areas of medical practice, such as nursing \citep{Corbin.1998}, and today it has found widespread adoption in medical practice \citep{Henly.2017,Larsen.2017,RoyalMarsden.2015}.

The central concept in the trajectory framework is the so-called trajectory \citep{Corbin.1988,Corbin.1991}, according to which a chronic disease undergoes different phases. These are listed in \Cref{tbl:trajectory_phases} as follows: (1)~an acute phase represents the most critical situation, often with a need for immediate treatment to control symptoms, (2)~an unstable phase entails a lack of full control over symptoms, and (3)~a stable phase is one in which symptoms are controlled. These trajectory phases (acute, unstable, stable) are managed by care providers and thus fall within the scope of this paper. Additional trajectory phases (\ie, pretrajectory, trajectory onset, crisis) refer to the time before diagnosis, while others are relevant for palliative care (\ie, downward phase and decease, which describe a potential death). Previous research has repeatedly validated the trajectory framework for a variety of chronic diseases. Examples include, for instance, stroke rehabilitation \citep{Burton.2000}, HIV/AIDS \citep{Corless.2000}, epilepsy \citep{Jacoby.2008}, and cancer \citep{Klimmek.2012}.

\begin{table}[H]
\caption{Trajectory Phases of Chronic Diseases \citep{Corbin.1991}.}
\scriptsize
\centering
\def\arraystretch{1.35}
\vspace{0.05cm}
\begin{tabular}{l p{12.5cm}}
\toprule
\textbf{Trajectory phase} & \textbf{Description} \\
\midrule
Acute & Severe complications; the aim is stabilizing the condition through (ambulant) hospitalization \\
Unstable & Course of disease and symptoms not fully controlled by regimen; continuing care but no further hospitalization \\
Stable & Course of disease and symptoms are controlled by treatment regimen; no need for hospitalization, but self-management practices are often introduced \\ 
\bottomrule
\multicolumn{2}{p{14cm}}{\emph{Note:} Additional trajectory phases apply to the time before diagnosis and to palliative care (beyond the scope of this paper).}
\end{tabular}
\label{tbl:trajectory_phases}
\end{table} 

For chronic disease management, the trajectory framework has important implications: the trajectory phases refer to different disease dynamics and thus demand different forms of care \citep{Larsen.2017}. Accordingly, patients in acute, unstable, and stable phases should essentially be treated as different cohorts. Hence, the decision-making problem for health professionals is to manage the trajectory phase by choosing a treatment regimen that is \textquote{specific to the illness phasing} \citep{Corbin.1991}. Put differently, if treatment regimen and trajectory do not match, the treatment is ineffective in managing the current disease dynamics, and thus represents over- or under-treatment \citep[\eg,][]{Kazemian.2019}. If the underlying trajectory phase changes, the treatment regimen must be adjusted accordingly. 

When operationalizing the trajectory framework in practice, health professionals usually follow a two-stage approach \citep{Larsen.2017}: First, the patient's symptoms are examined in order to identify the current trajectory phase. By knowing the current phase, healthcare professionals can adapt the treatment plan accordingly, that is, \emph{which} treatment regimen (acute, unstable, stable) to choose and \emph{when} to update the treatment regimen. Thereby, the trajectory framework ensures that the underlying disease dynamics (rather than merely symptoms) are treated. In a second stage, the design of the treatment regimen is chosen, \eg, among dimensions of the treatment regimen such as the type of medication and its dosage (cf. the next section for details). The design is a control task with the objective of maintaining the patient's present status of symptoms. In contrast, deciding upon a trajectory phase, and thus a corresponding treatment regimen, represents an identification task. That is, the course of the symptoms is interpreted in order to determine the trajectory phase \citep{Corbin.1991}. 

Applying the trajectory framework in practice is subject to challenges. First, the trajectory phases are usually recurrent and long-lasting, that is, spanning weeks or sometimes even months. Hence, close monitoring is needed, so that the treatment regimens can be adjusted dynamically. Second, it is left open how practitioners in healthcare and nursing can actually infer the current trajectory phase \citep{Corbin.1991}. Even though the past course of symptoms should be analyzed, clinical practice often lacks full information of this sort due to a lack of longitudinal monitoring. Third, trajectory phases and symptoms are only stochastically related. Within each phase, there may be temporary periods in which the symptoms can vary (\ie, temporary reversal or worsening of symptoms), although the underlying trajectory phase remains unchanged \citep{Corbin.1991}. Hence, this represents a source of human error \citep[\eg,][]{Burton.2000}, as the correct trajectory phases can often only be identified post-hoc. 

\subsection{Decision Support for Chronic Disease Management}
\label{sec:literature_decision_support}

In order to aid chronic disease management, prior literature has developed decision support to optimize treatment planning. This is summarized in the following. 

Risk scoring is supposed to foster informed decision-making and thus helps both patients and health professionals by forecasting the future course of a disease, such as expected health outcomes \citep[\eg,][]{Bertsimas.2017,Lin.2017,mueller2020longitudinal,fu2012risk} or readmission risk \citep[\eg,][]{Ayabakan.2016,Bardhan.2015}. Some works even derive explicit target levels for certain risk factors \citep{Helm.2015,Kazemian.2019}. The pathways of chronic diseases are described based on models that can capture their long-term and recurrent dynamics. Hence, common choices are simple Markov chains or semi-Markov models \citep[\eg,][]{Chou.2017,Srikanth.2015}. However, these models assume the trajectory phases to be known a~priori, whereas our objective is to identify them. Similarly, there have been a few works that accommodate hidden states, yet with clear differences from our work: These approaches operate on diagnosis codes \citep{Alaa.2018,Liu.2015b,Wang.2014b} or hospital readmissions \citep{Ayabakan.2016,Martino.2018}, but not symptoms. Furthermore, their purpose is predicting future risk, whereas we recover latent trajectory phases. As we shall see later, our objective requires a tailored modeling approach. 

Various approaches have been developed that help in determining the design of a predefined treatment regimen. This involves, for instance, the type of medication \citep[\eg,][]{Bertsimas.2017,Zargoush.2018} and the corresponding dosage \citep[\eg,][]{Ibrahim.2016,Lee.2018b,Murphy.2003,Negoescu.2017}. These works commonly build upon Markov decision processes \citep[cf.][for an overview in healthcare]{Schaefer.2005}, whereby, for instance, diagnoses represent the input and actions are in the form of, \eg, dosage. Due to chronicity, the objective is usually not to cure but rather to stabilize health outcomes. In some cases, the models have been extended by latent dynamics (\ie, partially-observable Markov decision processes), which can directly account for the unobservable responsiveness of individuals to medication \citep{Ibrahim.2016,Negoescu.2017}. Similarly, other works model the timing of specific interventions, \eg, hepatitis C drugs \citep{Liu.2015}, antiretroviral therapy \citep{Shechter.2008}, dialysis \citep{Lee.2008}, or liver donations \citep{Alagoz.2004}. However, all of the previous works concern the optimal control of a single treatment regimen that has already been identified, rather than managing the trajectory of chronic diseases across multiple treatment regimens.  

The above works study chronic disease management whereby health professionals manage the design of a single treatment regimen and, in this context, the treatment regimen is assumed to be known and fixed. As such, these works do not aid health professionals in managing the course of chronic diseases across multiple treatment regimens. However, this is necessary in practice: chronic diseases are characterized by a long-lasting progression that undergoes acute, unstable, and stable trajectory phases (\ie, as defined in the trajectory framework). Each of these phases represents a different patient cohort and, because of this, regular updates to the treatment regimen are necessary. Hence, managing the course of chronic diseases requires a model that identifies the current trajectory phases such that a treatment regimen for acute, unstable, or stable phases is recommended. 

\subsection{Hidden Markov Models for Management Decision-Making}

Hidden Markov models represent a flexible class of models with latent dynamics, whereby the time series undergoes transitions between a discrete set of unobservable states \citep{MacDonald.1997,Rabiner.1989}. This formalization has found widespread application in management decision-making \citep[\eg,][]{dong2007hidden,jiang2016hidden}. Examples include finance/insurance \citep[\eg,][]{reus2016dynamic,avanzi2021modelling}, engineering \citep{zhou2010model}, marketing \citep[\eg,][]{HattWWW,MKSC,Netzer.2008,ClickstreamDMM}, or out-of-stock prediction \citep{Montoya.2019}. 

The HMM-based framework has several benefits: First, it recovers the latent states which are linked to managerially relevant interpretations. Second, certain states are usually associated with management interventions. For instance, in the aforementioned works, the latent states encode the latent activity level of customers (\eg, loyalty, intrinsic motivation), which are then used for timing interventions. Following this motivation, the HMM-based framework appears promising for chronic disease management, where latent states can be used to suggest intervention points for treatment planning.

% HMMs in healthcare

Hidden Markov models have previously been applied in disease modeling, albeit for a different purpose than in our work. For instance, other works have used HMMs to study addictive behavior \citep[\eg,][]{DeSantis.2011,Shirley.2010}, comorbidities \citep{Maag}, critical conditions in intensive care units \citep{AttDMM}, organ failure \citep{Bartolomeo.2011,Martino.2018}, mental illnesses \citep[\eg,][]{Scott.2005}, or telehealth \citep{Ayabakan.2016}. In contrast to that, we propose a novel model (\ie, our \model) and adapt it to treatment planning for chronic diseases. 

\subsection{Extensions of the Na{\"i}ve HMM}

Motivated by chronic disease management, we later extend the na{\"i}ve HMM \citep{MacDonald.1997,Rabiner.1989}  to clinical settings in three ways: (1)~a variable-duration component, (2)~a copula approach, and (3)~heterogeneity in the transitions. Previously, several works have used \emph{either} variable-duration HMMs \citep[\eg,][]{Barbu.2008,Chiappa.2014,Kundu.1998,Limnios.2008,Murphy.2002,MKSC,Yu.2010} \emph{or} copula HMMs \citep[\eg,][]{CopulaHMM1,CopulaHMM2}. However, we are not aware of a HMM combining both variable-duration \emph{and} copulas. This later gives rise to a novel \model.

In HMMs, the variable-duration component changes the transitions. Recall that, in a na{\"i}ve HMM, transitions are Markovian; that is, they can only depend on the single previous state. However, such behavior opposes our theoretical knowledge about chronic diseases for which dynamics depend on the past health trajectory \citep[\eg,][]{Bakal.2014} and thus demand a variable-duration component. By contrast, in a variable-duration HMM, transitions are semi-Markovian \citep{Yu.2010}; that is, they are allowed to additionally depend on the duration of the previous latent state and thus become non-stationary. Example applications have been, \eg, in handwriting recognition, marketing, or DNA analysis \citep{Kundu.1998,Limnios.2008,MKSC}. 

Copula are widely used to model the dependence structure among multivariate observations \citep[\eg, in engineering; see][]{Brunel.2005,Brunel.2010}. A vast stream of literature has used copulas but outside of HMMs. There are also some copula HMMs \citep[\eg,][]{CopulaHMM1,CopulaHMM2}. Later, the copula approach is desirable for our research as allows the model to reflect that some symptoms are likely to co-occur \citep{Martino.2018}. 

In sum, variable-duration HMMs and copula HMMs have been developed separately. Building upon that, we later combine both into a novel model, \ie, a so-called variable-duration copula HMM (named \model).

\section{Model Development}
\label{sec:method}

An overview of key notation is provided in \Cref{tbl:notation}.

\begin{table}[H]
\caption{Notation.}
\scriptsize
\centering
\vspace{0.05cm}
\begin{tabular}{l p{12cm}}
\toprule
\textbf{Symbol} & \textbf{Description}\\
\midrule
$i$ & Patient index with $i = 1, \ldots, N$ \\
$t$ & Time step with $t = 1, \ldots, T$ \\
$m$ & Index enumerating different health measurements \\
$Y_{it} \in \mathbb{R}^m$ & Health measurements (multivariate) of patient $i$ at time $t$\\
$Y_{it}^{(1)}, \ldots, Y_{it}^{(m)}$ & Elements in $Y_{it}$ \\
$S_{it}$ & Latent state for patient $i$ at time $t$, capturing trajectory phase \\
$\varsigma$ & Number of states (later: $\varsigma = 3$ corresponding to acute, unstable, stable) \\
$\gamma_{it}^{j\to k}$ & Transition probability of patient $i$ to move from state $j$ at time $t$ to state $k$ at time $t+1$ \\
$\Gamma_{it}$ & Transition probability matrix with elements $\gamma_{it}^{j\to k}$ \\
$b_s(Y_{it} = y)$ & Emission function given the likelihood of observing a health measurement $y$ \\
$b_s^{(1)}, \ldots, b_s^{(m)}$ & Marginal emission functions of $b_s$ \\
$d_{it}(j)$ & Duration of patient $i$ in spending in latent state $j$ at time $t$ \\
$\delta_{jl}$, $\omega_{jl}$, $\beta_{jl}$ & Coefficients in multinomial logit function specifying the transitions \\
$L$ & Likelihood \\
$C_s$ & Copula function parameterized by $s$ (or, alternatively, $C$) \\
$F$ & Cumulative distribution function (with marginal cumulative distribution functions $F_1, \ldots, F_m$) \\
$u,v$ & Variables $\in [0, 1]$ \\
$\nu_s$, $\nu'_s$ & Parameters in copula (\eg, $\nu_s$ is the tail dependence in survival Gumbel copula) \\
$S_{it}^\ast$ & Recovered latent state via forward algorithm under the most likeliest latent state sequence \\
$\phi_{it}$ & Recommended treatment regimen $\phi_{it} \in \{ \text{acute}, \text{unstable}, \text{stable} \}$ \\
\bottomrule
\end{tabular}
\label{tbl:notation}
\end{table} 

\subsection{Problem Statement}

We now translate the trajectory framework into a data-driven dynamic model for personalized treatment planning. The input is given by symptoms and patient risk profiles in longitudinal form. Based on these, we model the progression of a chronic disease through acute, unstable, and stable trajectory phases. The trajectory phases are unobservable and, as a remedy, we model these as latent states. This follows prior research \citep{Corbin.1991} according to which the relationship between symptoms and trajectory phases should be considered to be stochastic. Then the objective of our model is to recover the latent states in order to yield dynamic suggestions for treatment regimens. 

The above decision problem is formalized via a tailored variable-duration copula hidden Markov model. The HMM-based framework \citep[\eg,][]{JMIR} allows us to formalize the progression of a disease over time, while modeling trajectory phases as latent states that can be dynamically inferred. In our proposed \model, we extend the na{\"i}ve HMM \citep{MacDonald.1997,Rabiner.1989}  in three ways: (1)~a variable-duration component, (2)~a copula approach for multivariate emissions, and (3)~heterogeneity in the transition. These are motivated in the following.

% VD

For (1), a variable-duration component is integrated into the model in order to account for how long a patient experienced a trajectory phase. The variable-duration component \citep{Yu.2010} changes the latent dynamics to become semi-Markovian, which supersedes the na{\"i}ve HMM where latent dynamics are only Markovian (as transitions can only depend on the single previous state). Yet the latter is not applicable when modeling the trajectory of chronic diseases \citep[\eg,][]{Chou.2017}: A longer duration in a stable trajectory phase means that the condition has likely subsided and that the future trajectory will remain in a stable phase. Analogously, longer exposure to an acute trajectory phase negatively afflicts the overall health condition and should thus increase the probability that the next trajectory phase will also be acute \citep{Bakal.2014}. Hence, our \model considers additionally the duration of being in a latent state (rather than just the latent state itself). The duration is itself latent and, to model this, a variable-duration component must be used. 

For (2), a copula approach is used in order to yield multivariate emissions and thus model multiple symptoms. This is demanded by medical research \citep[\eg,][]{Jensen.2015}, where the severity of chronic diseases is monitored along two (or more) health measurements that are usually highly interdependent. For instance, in the case of low back pain, a high pain intensity usually coincides with a severe limitation of activity and vice versa. Therefore, our \model builds upon a dependence structure among health measurements in the form of a copula approach. 

For (3), additional risk factors are integrated in our \model. This considers that the progression of chronic diseases entails considerable between-patient heterogeneity \citep[\eg,][]{Ayvaci.2017,Ayvaci.2018,Bardhan.2015,Lin.2017}, and, hence, we control for patient-specific risk factors. In our model, risk factors affect the propensity with which a patient transitions between latent states. 

\subsection{Proposed Variable-Duration Copula HMM}

The na{\"i}ve HMM consists of four components, namely, (1)~the observations, (2)~the latent states, (3)~a transition component, and (4)~an emission component linking states and observations \citep{MacDonald.1997,Rabiner.1989}. Observations are given by health measurements (\eg, symptoms), whereas the latent states are not observable (these reflect the different trajectory phases). 

\begin{figure}[H]
\OneAndAHalfSpacedXI
\centering
\includegraphics[width=0.7\textwidth]{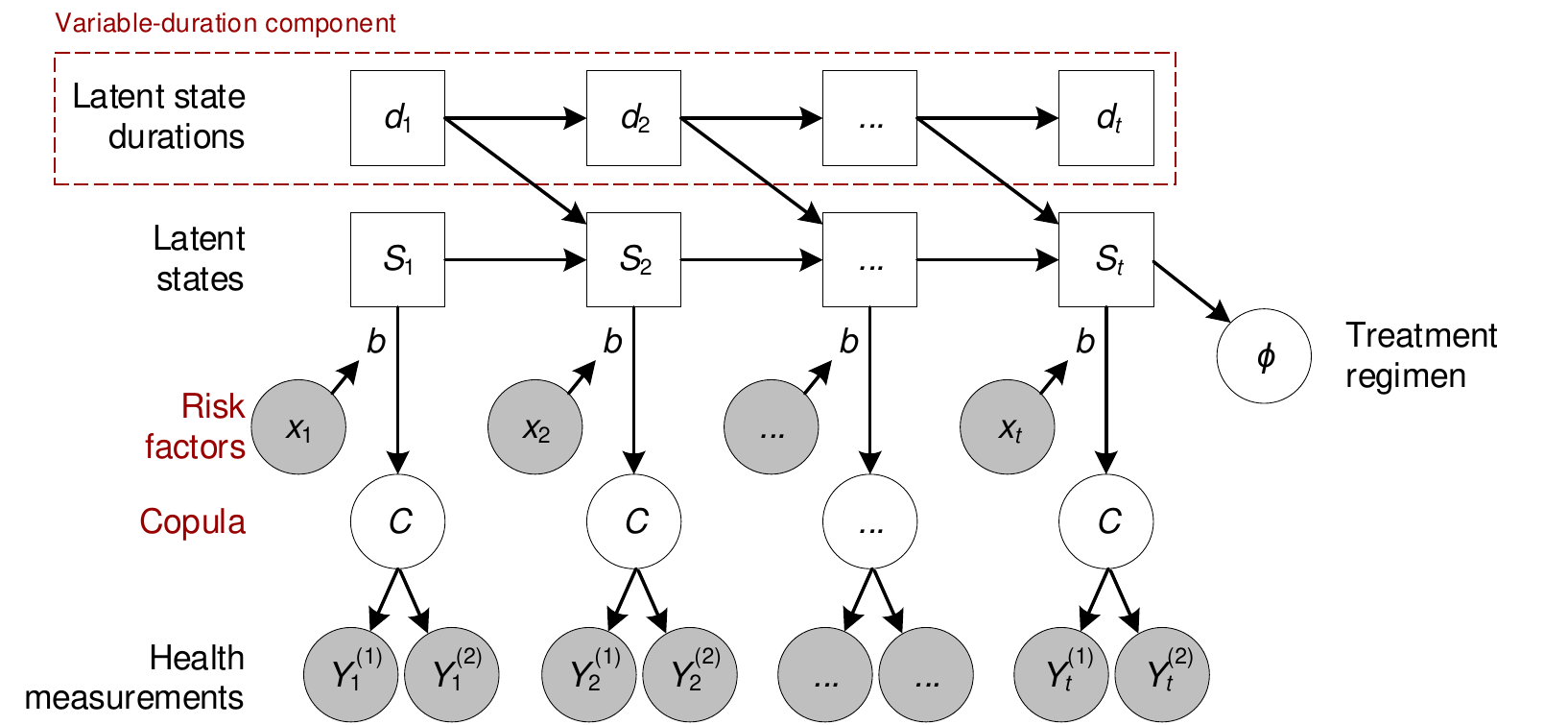}
\label{fig:schema}
\caption{Proposed variable-duration copula hidden Markov model. Shown is the \model for a given patient $i$ (subscript omitted for better readability). Health measurements are observable (gray shading), whereas the latent states are not (these reflect the different trajectory phases).}
\end{figure}

The four components (1)--(4) from above are adapted as follows to obtain our \model:
\begin{quote}
\begin{enumerate}
\item \emph{\textbf{Observations.}} The input is given by a multivariate sequence of $m$ health measurements over time (\ie, symptoms such as pain or activity limitation). Formally, we refer to the health measurements by $Y_{it}\in\mathbb{R}^m$ for patients $i = 1, \ldots, N$ and time steps $t = 1, \ldots, T$. We denote the $m$ elements in the vector $Y_{it}$ by $Y_{it}^{(1)}, \ldots, Y_{it}^{(m)}$. 
\item \emph{\textbf{Latent states.}} The latent states reflect the different trajectory phases. For each patient $i$, our \model assumes a latent state sequence $S_{it}\in\{1,\ldots,\varsigma\}$. By definition, the latent state sequence (\ie, the trajectory phases) cannot be observed directly; instead, it must be retrieved from the observations $Y_{it}$. The number of latent states is later determined as part of our evaluation, where we confirm that the best fit is yielded by $\varsigma = 3$ states (\ie, acute, unstable, and stable), analogous to the trajectory framework \citep{Corbin.1991}. 
\item \emph{\textbf{Transitions.}} The transition between the latent states follows a stochastic process that is defined by a transition probability matrix $\Gamma_{it}\in[0,1]^{\varsigma \times \varsigma}$. Here the probability of a patient $i$ moving from state $j$ at time $t$ to state $k$ at time $t+1$ is given by $\gamma_{it}^{j\to k} = P(S_{i,t+1} = k \mid S_{it} = j)\in[0,1]$. The transitions must fulfill $\sum_{k=1}^{\varsigma}\gamma_{it}^{j\to k} = 1 $ for each state $j = 1, \ldots, \varsigma$, patient $i$, and time $t$. The transition probabilities further account for between-patient heterogeneity that is provided by $x_{it}$ (\eg, treatments, patient-specific risk factors such as gender or age). 

Mathematically, the transitions in the na{\"i}ve HMM are constrained by the Markov property \citep{Murphy.2012}. That is, for a sequence of latent states $S_{i1}, \ldots, S_{it}$, the next state $S_{it}$ can only depend on the previous state $S_{i,t-1}$ and no other previous state (and also not on its duration), \ie,  $P(S_{it} \mid S_{i1}, \ldots, S_{i,t-1}) = P(S_{it} \mid S_{i,t-1})$. Hence, all previous states $S_{i1}, \ldots, S_{i,t-2}$ (and their durations) are neglected. 
\item \emph{\textbf{Emissions.}} The emission probability defines the likelihood of observing a certain symptom given the current latent state. This introduces another stochastic process that models state-dependent observations. Given patient $i$, the emission probability of $Y_{it}$ is given by $b_{s}(Y_{it}=y) = P(Y_{it}=y\mid S_{it} = s)$ with latent state $S_{it}=s$ at time step $t$. Note that our model operates on multivariate observations and, hence, the variables $Y_{it}$ are $m$-dimensional. Formally, we later refer to different marginal emission functions as $b_s^{(1)}, \ldots, b_s^{(m)}$. In our \model, their dependence structure is modeled as part of the copula approach in \Cref{sec:copula_model}.\\

\noindent
In our study, we observe multiple measurements which are given on a discrete scale~(\ie, Likert scale). Consistent with prior research \citep{Goulet.2017}, we thus model the marginal distributions as truncated Poisson distributions. For $k=1,\ldots,m$, the marginal emission is given by
{\small\begin{equation}
b_s^{(k)}(Y_{it}^{(k)} = y_k) = P(Y_{it}^{(k)} = y_k\mid S_{it} = s) = \frac{{\lambda^{(k)}_s}^{y_k}e^{-\lambda^{(k)}_s}}{y_k!}\times R_s^{(k)}.
\end{equation}}
Here the parameter $\lambda_s^{(k)}$ depends on both the margin $k$ as well as the latent state $s$. The term $R_s^{(k)}$ is a normalization factor resulting from the truncation at the maximum of the corresponding scale $L^{(k)}$ of the distribution and equals $R_s^{(k)} = \sum_{l=0}^{L^{(k)}} \frac{{\lambda^{(k)}_s}^{l}e^{-\lambda^{(k)}_s}}{l!}$.
\end{enumerate}
\end{quote}
\noindent
Different from a na{\"i}ve HMM, our \model model accommodates (1)~not only latent states but also their duration (as modeled by the variable-duration component), (2)~a copula to account for the dependence structure among health measurements, and (3)~additional patient-specific risk factors to consider between-patient heterogeneity. This is shown in \Cref{fig:schema}.

In the following, we provide details on the variable-duration component and the copula approach inside our \model. Both of which are tailored to the use of our proposed \model. Importantly, a custom likelihood is derived for this which is carefully tailored to our proposed variable-duration copula HMM. 

\subsection{Variable-Duration Component}

As part of the variable-duration component in our \model, we now specify the transition probabilities $\gamma_{it}^{j\to k}$ based on further controls $x_{it}$ (\eg, patient-individual risk factors) and the duration spent in a latent state. The latter is of particular importance: It allows us to consider the past latent state sequence. Thereby, we overcome limitations due to the Markov property that are inherent to na{\"i}ve HMMs \citep{Yu.2010}. Specifically, na{\"i}ve HMMs assume that transition probabilities are independent of time and they are therefore treated as stationary. However, as argued above, there is evidence in medical research that this assumption is too restrictive and, instead, our setting demands that the transition probability depends on the time spent in the current state. This is formalized in the variable-duration component of our \model.

We make use of the following notation. Formally, the transition probability $\gamma_{it}^{j\to k}$ defines the probability of patient $i$ moving at time $t$ from a latent state $j$ to a latent state $k$. To this end, let $d_{it}(j) $ denote the prior duration (in weeks) that patient $i$ spent in the latent state $j$ at a certain time $t$. It gives the consecutive duration since the last transition. Analogous to the latent state, the duration $d_{it}(j) \in\left\{1,\ldots,t\right\}$ is also latent. This prevents us from simply inserting $d_{it}(j)$ in the model, since it is not directly observable. Instead, one requires an approach where both latent states and latent state durations are modeled jointly, as in a variable-duration component.

The variable-duration component distinguishes two cases when modeling the transition probability $\gamma_{it}^{j\to k}$: (1)~The patient remains in the current latent state, \ie, $j = k$. We refer to this as a recurrent transition. This occurs with probability $\gamma_{it}^{j\to j}$. (2)~The latent state of the patient changes to a different state, \ie, $j \neq k$. We refer to this as a non-recurrent transition, which occurs with probability $1-\gamma_{it}^{j\to j}$. 

Both the recurrent and non-recurrent transitions are modeled in the variable-duration component, while further incorporating additional sources of heterogeneity $x_{it}$, such as risk factors or previous health measurements. As as suggested in previous works \citep[\eg,][]{Netzer.2008}, we model the risk factors inside the transitions via a multinomial logit function. The multinomial logit considers the current state as a base case and then compares it to the probability of moving to a different state. This allows the propensity of moving between states to differ based on patient characteristics. Formally, we specify
{\small\begin{equation}
\label{eq:transition_naive_vdhmm}
\gamma_{it}^{j\to k}=P\left(S_{i,t+1} = k\mid S_{i,t}=j, d_{it}(j)\right) = 
\begin{cases}
\cfrac{1}{1+\sum_{l\neq k} \exp(\delta_{jl} + \omega_{jl} d_{it}(j) +{x}_{it}^T \beta_{jl} )}, & \text{ for } k = j, \\	\cfrac{\exp(\delta_{jk} + \omega_{jk} d_{it}(j) + {x}_{it}^T \beta_{jk} )}{1 + \sum_{l\neq k} \exp(\delta_{jl} +\omega_{jl} d_{it}(j) +  {x}_{it}^T \beta_{jl})}, & \text{ for } k \neq j .
\end{cases}
\end{equation}}
with intercepts ${\delta}_{jl}$ and coefficients $\omega_{jl}$ and ${\beta}_{jl}$ for $j,l = 1, \ldots, \varsigma$. Consistent with \citet{Netzer.2008}, %, Shirley.2010},
 we set the parameters ${\beta}_{jj}$, ${\omega}_{jj}$ and ${\delta}_{jj}$ to zero to ensure identifiability of the parameters.\footnote{\SingleSpacedXI\footnotesize We also experimented with alternative specifications of the transition matrix, though with inferior results. Specifically, we tested different structural assumptions in the transition matrix, yet this proved not to be beneficial. For example, previous work has encoded a funnel structure in which transitions could occur only between neighboring states. However, this approach resulted in a model fit that was inferior.}  

The above variable-duration component has multiple implications. First, the dependence on the latent state duration $d_{it}$ essentially relaxes the Markov property that limits na{\"i}ve HMMs. Second, the transition probabilities are no longer time-homogeneous, as they depend on the prior duration in the current state. This holds true for both recurrent and non-recurrent transitions. Third, the variable-duration component allows states to become more \textquote{sticky} with $d_{it}(j)$. This is captured via the parameters $\omega_{jl}$ for $j\neq l$. If $\omega_{jl}$ is positive, the probability of a non-recurrent transition increases with a longer duration $d_{it}(j)$. If $\omega_{jl}$ is negative, a recurrent transition becomes more likely. Fourth, our specification benefits from direct interpretability, as we can identify the extent to which risk factors influence disease dynamics. Fifth, if $\omega_{jl}$ are all held at zero, the na{\"i}ve HMM becomes a special case of our \model. Because of this, the likelihood of the HMM differs from that of our \model and, in order to estimate our model, must be derived.

Our \model model is estimated via Markov chain Monte Carlo~(MCMC) sampling by deriving the likelihood 
{\small\begin{equation}\label{eq:ll_old}
L=\prod_{i=1}^{N}P\left(y_{i1},\ldots,y_{iT}\right) = \prod_{i=1}^{N}\sum_{s_1=1}^{\varsigma}\sum_{s_2=1}^{\varsigma}\cdots\sum_{s_T=1}^{\varsigma}\left[P(S_{i1}=s_1)\times\prod_{t=2}^{T}\gamma_{it-1}^{s_{t-1}\to s_t}\times \prod_{t=1}^{T}b_{s_t}(Y_{it}=y_{it})\right].
\end{equation}}%
If we assume the margins of the observations $Y_{it}$ to be independent given the current latent state $S_{it}$, the above equation simplifies to
{\small\begin{equation}\label{eq:ll_old_independence}
L= \prod_{i=1}^{N}\sum_{s_1=1}^{\varsigma}\sum_{s_2=1}^{\varsigma}\cdots\sum_{s_T=1}^{\varsigma}\left[P(S_{i1}=s_1)\times\prod_{t=2}^{T}\gamma_{it-1}^{s_{t-1}\to s_t}\times \prod_{t=1}^{T}\prod_{k=1}^m b_{s_t}^{(k)}(Y_{it}^{(k)}=y_{it}^{(k)})\right].
\end{equation}}
By sampling from the likelihood $L$, we can thus directly estimate $\omega_{jl}$ from the data. In practice, a direct evaluation of the likelihood is computationally intractable and we thus apply the forward algorithm for an efficient calculation \citep{Yu.2010}. All estimation details are reported in the supplements.

\subsection{Copula Approach for Modeling Dependence Structures}
\label{sec:copula_model}

\subsubsection{Overview.}

Health management commonly monitors the progression of diseases along multiple symptoms \citep[\eg,][]{Jensen.2015}; however, symptoms are usually not unrelated. Rather, they co-occur in a specific manner: (1)~either \emph{all} (or almost all) symptoms are absent when the patient has recovered, or (2)~the condition is indicated by \emph{some} -- but not necessarily all -- symptoms due to differences in how patients respond to a disease. For instance, patients with stable low back pain experience an absence of both pain and activity limitation, whereas acute low back pain is usually characterized by severe pain or activity limitation, though rarely both. In other words, the absence of one symptom makes it more likely that other symptoms will also be absent. Altogether, this results in a lower tail dependence among health measurements that must be modeled accordingly. 

In this section, we provide a rigorous approach for modeling the dependence structure among multivariate (discrete) emissions. The dependence structure represents a clear difference to the na{\"i}ve HMM where independence is assumed. Formally, in the above likelihood, independence was needed in order to rewrite $b_s(Y_{it}=y)$ in the likelihood as $\prod_{k=1}^{m} b_s^{(k)}(Y_{it}^{(k)}=y_k)$, \ie, when moving from \Cref{eq:ll_old} to \Cref{eq:ll_old_independence}. Hence, in order to consider a dependence structure, we now need to derive a new likelihood. To accomplish this, the likelihood of our \model is updated by proceeding in two steps. We first formalize the distribution function $b_s(Y_{it} \leq y)=P(Y_{it} \leq y\mid S_{it}=s)$ via a copula function $C_s$ operating on the marginal distribution functions. Here the copula function $C_s$ provides a flexible tool for modeling the nature of the dependence structure (\ie, a lower tail dependence). Second, we use the resulting distribution function to derive the new emission $b_s(Y_{it}=y)$ subject to a copula-based dependence structure. The new emission is then simply inserted into the likelihood from \Cref{eq:ll_old}, so that we estimate the model parameter by applying the previous MCMC sampling technique.

\subsubsection{Copula-Based Distribution Function.}

Let $C_{s}(\cdot, \ldots, \cdot)$ denote a copula function \citep{Joe.1997}. A copula is a generalized correlation function, which allows for a stronger dependence in certain parts of the distribution. Formally, it links the individual marginal distributions of an $m$-dimensional input to a joint cumulative distribution function, while introducing a desired dependence structure between the margins. Depending on which copula is chosen, $C_{s}$ might be parameterized by further variables that, for instance, control for the strength of the dependence. In our work, the parameters are modeled as state-specific, which is indicated in the copula by the subscript $s$. A state-specific $C_s$ is needed for our research, as there should be different dependencies across states (\eg, a stronger dependence of the tails in acute as opposed to stable states). The copula can be freely chosen; however, our setting demands a lower tail dependence due to the specific characteristics of health measurements and, hence, we use a survival Gumbel copula. This choice is discussed later.

Formally, a copula $C\colon[0,1]^m\to[0,1]$ is an $m$-dimensional cumulative distribution function. For every $m$-dimensional cumulative distribution $F\colon\mathbb{R}^m\to[0,1]$ with marginal cumulative distribution functions $F_1,\ldots,F_m$, the existence of a copula function, such that $F(x_1,\ldots,x_m) = C(F_1(x_1),\ldots,F(x_m))$, is guaranteed by Sklar's theorem \citep{Joe.1997}. If all margins are continuous, the copula $C$ is unique. However, the same does not hold if one of the margins is discrete \citep{Genest.2007}, as in our case.

We now use the copula to formalize a dependence structure between the marginal distribution of $b_s(Y_{it} \leq y)$. This is given by 
{\small\begin{align}
b_s(Y_{it} \leq y) & = P(Y_{it}^{(1)}\leq y_1,\ldots,Y_{it}^{(m)}\leq y_m\mid S_{it}=s) \\
& = C_{s}\left(P(Y_{it}^{(1)}\leq y_1\mid S_{it}=s),\ldots,P(Y_{it}^{(m)}\leq y_m\mid S_{it}=s)\right) .
\end{align}}%
Here the dependence between observations $y_1, \ldots, y_m$ is modeled by \textquote{linking} the separate marginal distribution functions $P(Y_{it}^{(1)}\leq y_1\mid S_{it}=s)$, \ldots, $P(Y_{it}^{(m)}\leq y_m\mid S_{it}=s)$ inside the copula \citep[\eg,][]{Nikoloulopoulos.2013}. Based on the distribution function $b_s(Y_{it} \leq y)$, we derive the new emission $b_s(Y_{it}=y)$ in the following. 

\subsubsection{Copula-Based Emissions.}

For a given copula, our objective is to derive a new emission $b_s(Y_{it}=y)$. The new emission is then integrated into the likelihood of our \model, which allows us to estimate the model under a given dependence structure.  

For discrete observations, as in our study, we derive the new emission via
{\small\begin{align}
b_s(Y_{it} = y) &=P(Y_{it}^{(1)} = y_1,\ldots,Y_{it}^{(m)} = y_m\mid S_{it} = s)\\
&= \sum_{i_1=0}^1\cdots\sum_{i_m=0}^1(-1)^{i_1+\ldots+i_m} \, P(Y_{it}^{(1)}\leq y_1-i_1,\ldots,Y_{it}^{(m)}\leq y_m-i_m\mid S_{it}=s) \label{eqn:emission_derivation_sum} \\
&=\sum_{i_1=0}^1\cdots\sum_{i_m=0}^1(-1)^{i_1+\ldots+i_m} \, C_{\nu_s}\left(P(Y_{it}^{(1)}\leq y_1-i_1\mid S_{it}=s),\ldots,P(Y_{it}^{(m)}\leq y_m-i_m\mid S_{it}=s)\right) . \label{eqn:emission_derivation_copula}
\end{align}}%
Here \Cref{eqn:emission_derivation_sum} follows from the fact that the marginal likelihood can be derived from the corresponding distribution function as $b_s^{(k)}(Y_{it}^{(k)}=y_k) = P(Y_{it}^{(k)}\leq y_k\mid S_{it}=s) - P(Y_{it}^{(k)}\leq y_k\mid S_{it}=s)$, $k = 1, \ldots, m$. \Cref{eqn:emission_derivation_copula} is simply the result of inserting the distribution function from above. For continuous or mixed observations, the derivation can be readily extended \citep[see][]{Onken.2016}. 

The resulting model is estimated as follows. The new emission $b_s(Y_{it} = y)$ is used to update the likelihood from our \model. That is, we simply replace $b_{s}(Y_{it}=y)$ inside \Cref{eq:ll_old}, thereby yielding the new likelihood function for our \model. Based on the likelihood, the model parameters can be simply obtained via MCMC sampling, as detailed in the supplements.

\subsubsection{Choice of Copula Function.}

Our copula $C_s$ should accommodate tail dependence, so that an absence of symptoms appears jointly. In order to model this behavior, we draw upon a survival Gumbel copula. For $u,v\in[0,1]$, it is given by
{\small\begin{equation}
C_s(u, v) = u + v - 1 + \exp\left[ -\left( (-\log(1-u))^{\nu_s} + ( -\log(1-v))^{\nu_s} \right)^{\frac{1}{\nu_s}}\right] ,
\end{equation}}
where the parameter $\nu_s \geq 1$ controls the strength of the tail dependence. It can be shown that the survival Gumbel copula has positive lower tail dependence for $\nu_s>1$ and zero upper tail dependence for all $\nu_s$ \citep[see, \eg, ][]{Joe.1997}. For the special case of $\nu_s = 1$, the survival Gumbel copula reduces to independent observations.

For comparison, we also later experiment with other dependence structures (see supplements). On the one hand, we use no copula. This ensures independence. On the other hand, we use the Ali-Mikhail-Haq copula in order to test for a symmetric dependence structure. It is given by $C_s(u, v) = u\, v\, (1 - \nu_s' (1 - u) (1 - v))^{-1}$ with an additional parameter $\nu_{s}'\in [-1, 1)$. Here the symmetric dependence structure follows from its multiplicative form. However, our numerical experiments later confirm that the Gumbel copula is superior. 

Employing a copula approach has multiple benefits. First, it is highly flexible as different copula functions for modeling dependence structure can be chosen. The copula functions can even be parameterized. Second, a copula approach requires fairly little a priori knowledge, as parameters are estimated from data. Third, using a copula approach generalizes to both continuous and, in particular, discrete observations. The latter prohibits the use of alternative approaches from the literature, where both marginal distributions and observations must originate from the same family of distributions. 

\subsection{Inferring Treatment Regimens}

Our \model is used to recommend treatment regimens in a two-step procedure. First, we compute the latent states specific to a patient profile. Formally, we recover the latent state sequence $S_{it}$ from the observed health measurements $Y_{i1},\ldots,Y_{it}$ conditional on a given risk profile $x_{i1}, \ldots, x_{it}$. Second, the recovered latent state is mapped onto one of the three treatment regimens $\phi_{it} \in \{ \text{acute}, \text{unstable}, \text{stable} \}$. To this end, the recommended treatment regimen is determined via
\begin{align}
& \argmax\limits_{\phi_{it}}\; P( \phi_{it} \mid Y_{i1},\ldots,Y_{it},x_{i1},\ldots,x_{it} )  \\
= & \argmax\limits_{\phi_{it}}\; = P( \phi_{it} \mid S_{it}^\ast) \quad \text{with } \quad S^\ast_{it} = \argmax_{s} P(S_{it} = s \mid Y_{i1},\ldots,Y_{it},x_{i1},\ldots,x_{it}) 
\label{eqn:regimen2}
\end{align}
using the observed frequencies of ground-truth labels $\phi_{it}$ from the training data. Note that the patient-specific risk profile $x_{it}$ (\eg, past treatments, age, gender) is explicitly considered in the latent dynamics of our \model, and, owing to this fact, the predictions are personalized to a patient's risk profile. For instance, a patient with a high body mass index has a transition mechanism different from that of a patient with a low body mass index, because of which the latent state sequence and final prediction are also different.

We emphasize that the previous procedure is inherently dynamic; it considers the inputs only from time step 1 until a given $t$. Once a new health measurements becomes available, the previous procedure must be repeated. The predictions are then compared against the expert labels, which were, however, obtained with post-hoc knowledge. That is, the experts have made their annotation by considering the progression until $T > t$ and thus beyond the time step $t$. Altogether, this yields the corresponding framework for a dynamic ``on-line'' evaluation:  
\begin{quote}\begin{quote}
\begin{algorithm}[H]
\DontPrintSemicolon
\SingleSpacedXI
\footnotesize
\For{$t = 1,\ldots $}{
Receive new health measurement $Y_{it}$ \;
Compute $S_{it}^\ast = P(S_{it} = s \mid Y_{i1},\ldots,Y_{it},x_{i1},\ldots,x_{it})$ for $s=1,\ldots,\varsigma$ using the forward algorithm\;
Determine treatment regimen $\phi_{it}$ \; 
Compare $\phi_{it}$ against post-hoc annotation from experts
}
\end{algorithm}
\end{quote}\end{quote}

% step 

\vspace{0.3cm}
\noindent
Formally, we recover patient-specific latent states by determining the likeliest state $S_{it}$ conditional on $Y_{i1},\ldots,Y_{it}$. The latent state depends on the latent state duration, which again depends on multiple previous latent states, thus driving the overall computational complexity. An efficient computation scheme is is obtained via the forward algorithm \citep{Yu.2010}.

\section{Longitudinal Study of Low Back Pain}
\label{sec:study_setting}

Our empirical model for personalized treatment planning is validated based on a longitudinal cohort study of patients with non-specific low back pain. Non-specific low back pain represents a chronic condition for many subjects, who experience a recurrent, variable, and often long-lasting course. Low back pain is globally responsible for the greatest number of years lived with disability \citep{GBD.2017} and thus burdens both patients and society with extensive costs, while its causes are widely unknown \citep{Foster.2018,Hartvigsen.2018}. Health researchers comprised part of the authorial team and ensured that both the above model development and the following evaluation fulfill clinical standards. 

\subsection{Study Design}

This work builds upon an extensive, longitudinal, 52-week study of 928 patients. The actual design of the study was developed and undertaken by experts from the medical domain in a manner that adheres to common conventions and regulations of clinical studies \citep{Kongsted.2015,Nielsen.2016,JCE}. The data was obtained based on a prospective observational cohort study. The clinical study was conducted between September 2010 to January 2012 in Southern Denmark. Eventually, the duration of the study was set at one year, as this exceeds the usual length of low back pain episodes by several times \citep{Kongsted.2015}. Hence, for most patients, a length of 52 weeks is sufficient to observe multiple episodes with severe symptoms. All general practitioners in that region were invited to participate and recruit participants. To collect longitudinal data, an IT-based health monitoring system was used to send weekly text messages to patients asking for their pain level and activity limitations, and the patients responded via text message. The use of this IT-enabled system for health tracking allowed the study to scale to both a large number of patients and long study period

The cohort fulfilled the following criteria: between 18--65 years of age, possessed a mobile phone, could read sufficiently well to answer the survey questions, and were not pregnant or suspected of having a serious pathology that required referral for acute surgical assessment. A healthcare professional had diagnosed the subjects with non-specific low back pain based on each patient's history and a physical examination. 

The study consists of three components as follows. (1)~An extensive upfront survey collected information on various risk factors that describe the between-patient heterogeneity. These variables are detailed later. We use the term \textquote{risk factor}, while, from a medical perspective, these variables are not describing the onset but rather the course of a disease.\footnote{\SingleSpacedXI\footnotesize Consistent with other research in management science, we use the term \textquote{risk factor} to refer to variables linked to health outcomes. However, the healthcare experts from our authorial team prefer the term \textquote{prognostic factor} in the specific case of low back pain as it betters relates to the course of the condition (as opposed to risk factors that relate to the onset of the condition).} (2)~The longitudinal progression was monitored over the course of 52~weeks. Patients were asked to report their weekly levels of experienced pain and activity limitation. (3)~The longitudinal data was subsequently annotated by three medical experts, physicians with experience in treating low back pain who were also trained in the trajectory framework. For each patient and week, the medical experts then stated whether they would recommend an acute, unstable, or stable treatment regimen. The annotations from the medical experts were aggregated by the majority vote. Kendall's tau amounts to \num{0.89} and, hence, the annotations revealed considerable between-annotator agreement.\footnote{\SingleSpacedXI\footnotesize The annotations are discrete (\ie, acute, unstable, stable) but ordered. Hence, we must account for the ordinal nature, and measure the between-annotator agreement through a rank correlation coefficient, that is, Kendall's tau \citep{Kendall}.} The annotations are considered the ground truth for all subsequent evaluations. We emphasize that the medical experts had unique post-hoc knowledge due to our longitudinal dataset when they inspected trajectories, whereas our decision rules are later evaluated in an on-line manner to reflect the partial information (until a certain time step $t \leq T$) that would be available to health professionals in practice.

\subsection{Data Description}
\label{sec:data_sec}

The course of low back pain was monitored using two health measurements $Y_{it}$: (1)~The average pain intensity, which was collected each week on the so-called NRS scale \citep{Breivik.2008}. This represents the default procedure for a valid and reliable measurement of pain in clinical research. It measures pain on a scale from 0 (\ie, no pain) to 10 (\ie, worst pain imaginable). (2)~Activity limitation was measured based on the number of days characterized by activity limitation during the previous week and thus ranges from 0 to 7 days. 

The variable $x_{it}$ is used to describe the between-patient heterogeneity. It is obtained by appending the following covariates. First, treatments are likely to be associated with the progression of a disease. Here we follow \citet{Yan.2014} and control for the fact that interventions can be linked to state transitions. For low back pain, we expect treatments (\eg, massage, acupuncture) to affect the latent state within a few days and thus fed the variable into the model without time lag (however, this may be different for other chronic diseases). Second, we include a set of nine further patient-specific risk factors (see \Cref{tbl:list_risk_factors}): age, height, BMI, general health, gender, leg pain, duration of current pain episode, number of previous episodes, and physical work load. These risk factors have been carefully chosen based on clinical considerations (\ie, they entail a high prognostic capacity with regard to the future course of low back pain). Of note, duration of current pain episode and number of previous episodes should not be confused with the phase (acute, unstable, stable), which is a latent variable. Third, we include the previous health measurements $Y_{i,t-1}$. Such a first-order structure is informed by prior theory in medical research on modeling intrapersonal pain variability \citep{Mun.2019}.

Out of the original 928 patients in our study, we excluded 76 due to their failure to participate in the weekly monitoring, as well as 5 patients who reported the same pain intensity in each of the 52 weeks. This yields a final sample of 847 distinct patients. All of our models were trained on a random subset of \num{600} patients, while the remaining \num{247} samples represent our test set in order to evaluate the accuracy of treatment regimens that were correctly identified.

\subsection{Summary Statistics}

\Cref{tbl:list_risk_factors} lists summary statistics for our risk factors. For instance, in our study, \SI{62.55}{\percent} of all low back pain episodes lasted up to 2~weeks, \SI{13.69}{\percent} from 2--4 weeks, and \SI{10.68}{\percent} from 1--3 months. 

The course of low back pain is characterized by considerable variability. Pain sometimes vanishes temporarily, but often persists in the long run. Furthermore, we regularly note jumps from a pain intensity of zero to higher values (and vice versa). Examples are presented in \Cref{fig:problem_statement}. As a consequence, we also observe that the annotations from our medical experts are highly variable (\ie, each trajectory was annotated, on average, with more than two different treatment regimens). 

Both health measurements (pain level and activity limitation) are highly correlated (correlation coefficient of \num{0.65} as measured by Kendall's $\tau_b$).\footnote{\SingleSpacedXI\footnotesize The health measurements are collected on a discrete, numerical scale with ordering. Hence, to account for the ordinal nature, we must compute a rank correlation coefficient \citep{Kendall}.} Moreover, low values of pain intensity are reported considerably more often by patients then severe ones. For instance, \SI{57.0}{\percent} of all reported pain levels correspond to a rating of zero (\ie, no pain). This results in a mean value of \num{1.45} with a standard deviation of \num{2.21}. The activity limitation follows analogously. Here the mean value amounts to \num{0.79} with a standard deviation of \num{1.85}. However, in terms of activity limitation, we observe a substantial number of patients reporting the maximum value of $7$. Severe activity limitation is frequent even when the pain intensity is moderate or low.

\begin{table}[htb]
\SingleSpacedXI
	\centering
	\tiny
	\caption{Summary Statistics of Risk Factors.}	
	\begin{tabular}{llr}
		\toprule
		\textbf{Continuous risk factors}  & Statistic & Value \\
		\midrule
		Age (in years) & Range &18--65\\
		&Mean & 43.20 \\
		&Std. dev.& 11.44 \\
		\midrule
		Height (in cm) & Range & 153--201 \\
		&Mean& 175.99 \\
		&Std. dev.& 8.86 \\
		\midrule
		Body mass index (BMI) & Range & 18--59 \\
		&Mean & 26.26 \\
		&Std. dev.& 4.66\\
		\midrule
		General health & Range & 0--100\\	
		(EQ-5D scale) &Mean & 67.57 \\
		&Std. dev.& 20.49 \\
		\toprule
		\textbf{Categorical risk factors} & Levels & Relative frequencies  \\
		\midrule
		Gender & Female & \SI{45.93}{\percent}\\
		& Male & \SI{54.07}{\percent}\\
		\midrule
		Severity of leg pain & No pain & \SI{41.98}{\percent}\\
		& Mild pain & \SI{33.95}{\percent}\\
		&Moderate-severe pain & \SI{24.07}{\percent}\\
		\midrule
		Duration of current episode &0--2 weeks & \SI{62.55}{\percent} \\
		&2--4 weeks &  \SI{13.69}{\percent}\\
		&1--3 months & \SI{10.68}{\percent} \\
		&More than 3 months & \SI{13.09}{\percent}\\
		\midrule
		Number of previous episodes & None & \SI{16.11}{\percent} \\
		&1--3 & \SI{34.62}{\percent}\\
		&More than 3 & \SI{49.28}{\percent} \\
		\midrule
		Physical work load & Sitting & \SI{24.03}{\percent} \\
		&Sitting and walking & \SI{35.74}{\percent}\\
		&Light physical work & \SI{19.93}{\percent}\\
		&Heavy physical work & \SI{20.30}{\percent}\\
		\toprule	
		\textbf{Longitudinal monitoring}  & Statistic & Value \\
		\midrule
		Pain intensity (NRS scale; 0--10) 
		&Mean & 1.45 \\
		&Std. dev.& 2.21 \\
		\midrule
		Activity limitation (in days; 0--7) 
		&Mean& 0.79 \\
		&Std. dev.& 1.85 \\
		\midrule
		Interventions (at week level; yes/no) & Yes & \SI{1.92}{\percent} \\
		& No & \SI{98.08}{\percent} \\
		\bottomrule
	\end{tabular}
	\label{tbl:list_risk_factors}
\end{table}

\subsection{Dependence Structure of Observations}

We finally investigate the dependence structure between pain intensity and activity limitation. The association between pain intensity and activity limitation amounts to \num{0.65} as measured by Kendall's $\tau_b$, which is statistically significant at the \SI{0.1}{\percent} threshold. \Cref{tbl:pain_joint_distribution} reports the relative frequency of pain intensities conditional on the activity limitation~(in \%). Thereby, we can investigate which values frequently appear together. We observe a strong dependence between both dimensions that becomes especially evident at the lower tails. For example, we generally observe low pain levels given a low activity limitation. However, when conditioning on the maximum activity limitation, we find a wide range of pain intensities (between $2$ and $8$).

\begin{table}[htbp]
	\centering
	\TABLE{Dependence Structure Between Health Measurements.\label{tbl:pain_joint_distribution}}
	{\scriptsize
		\begin{tabular}{r@{\hskip 0.5cm} ccccccccccc}
			\toprule
			&\multicolumn{10}{c}{Pain intensity}\\
			\cmidrule(lr){2-12}	
			Activity limitation & 0 & 1 & 2 & 3 & 4 & 5 & 6 & 7 & 8 & 9 & 10 \\
			\cmidrule(lr){2-12}
      \csname @@input\endcsname correlations_main % load file 			
%			\input{correlations_main.tex}
%			\cmidrule(lr){2-10}
%			\input{correlations_margin.tex}	
			\bottomrule	
		\end{tabular}
	}{\hspace{-0.3cm}\emph{Notes:} The table reports the relative frequency of pain intensities conditional on the activity limitation~(in \%). Thereby, we can investigate which values frequently appear together. We note a strong dependence between both dimensions that becomes especially evident at the lower tails. Rows are standardized to sum to one. All cells with values above $10$ are shaded in light gray; above 25 in dark gray.}	
\end{table}

\subsection{Current Practice in Managing Low Back Pain}

Low back pain is known for the recurrent nature of its course \citep{Foster.2018,Hartvigsen.2018}. This is also reflected in the treatment regimens that are conventionally practiced \citep{Foster.2018,Traeger.2019}. For example, in a stable treatment regimen, patients primarily receive a combination of pain self-management, education, and graded exercise therapy. As with other chronic diseases, the treatment regimens aim at stabilizing the current trajectory phase. As is common in clinical research, the effectiveness of the different models is compared by computing the quality-adjusted life years (QALYs). Here we use the disability weights from \citet{Hoy.2014}.\footnote{\SingleSpacedXI\footnotesize \citet{Hoy.2014}have estimated the disability weights (DW) for mild acute low pain ($\mathit{DW} = 0.040$) and for severe chronic low back pain (mean $\mathit{DW} = (0.366+0.374)/2 = 0.370$). Here, a DW of one means that a complete year with the disease is discounted, while a DW of zero means that the year is enjoyed by a patient with full quality of life. If a decision model identifies the correct treatment regimen, we set $\mathit{DW} = 0$, while, otherwise, the $\mathit{DW}$ from above is used. We then approximate QALY via $I \times (1-\mathit{DW}) \times L$, where $I$ is the incidence and $L$ is the time frame of our data. }

Providing care for low back pain entails the following costs \citep{Whitehurst.2012}: A stable treatment regimen was estimated to have an annual costs of only USD~208.57 (GBP~\num{160.44}) per patient. Unstable and acute treatment regimens often involve physical therapies, and their yearly costs per patient are thus considerably larger, amounting to USD~374.14 (GBP~\num{287.80}) and USD~464.71 (GBP~\num{357.47}), respectively. Costs associated with managing low back pain usually accumulate over an extensive period, as this disease is known for causing the largest number of years lived with disability \citep{GBD.2017}. 

Choosing a treatment regimen that does not match the trajectory phase is considered ineffective when following the clinical guidelines \citep{Larsen.2017}; it fails to adapt to the underlying disease dynamics as it targets patients from a different cohort. For instance, pain self-management is highly effective in a stable trajectory phase, yet can barely maintain the course during an unstable or acute phase. Similarly, medication for acute treatment regimens should be avoided for less severe regimens due to strong side-effects \citep{Traeger.2019}. Analogous to \citet{Kazemian.2019}, we base our evaluations on the following cost structure, where costs are accumulated over all misidentified treatment regimens (\ie, the absolute deviation between suggested and needed treatment regimen, \ie between the used and the post-hoc required treatment costs treatment costs). To this end, more comprehensive treatment regimens than necessary are counted as over-treatment, while less comprehensive treatment regimens than necessary are counted as under-treatment.

\subsection{Baselines}
\label{sec:baselines}

As for all chronic diseases, health professionals are advised to consider the past course of the disease in their decision-making. Oftentimes, however, this information is not readily available due to a lack of longitudinal monitoring. To formalize decision-making in current practice, we consider a set of decision rules that vary in terms of the extent to which the past course is considered. As in earlier research \citep[\eg,][]{DAmato.2000,Ibrahim.2016}, we develop different myopic decision rules and extend them with data from the patient history (ordered by increasing complexity): 
\begin{enumerate}
\item {Myopic decision rule (symptoms only)}. Analogous to \citet{Ibrahim.2016}, we develop a myopic decision rule. Here we use a generalized linear model to predict the annotated regimen from the current health measurements $Y_{it}$ (and a pain-disability interaction term). 
\item {Myopic decision rule (risk factors)}. In order to accommodate between-patient heterogeneity, this model incorporates both $Y_{it}$ (with interaction term) and risk factors $x_{it}$  in a generalized linear model. 
\item {Dynamic decision rule (moving average)}. Here we extend the previous decision rule, so that it is additionally fed with the past history of health measurements via a moving average, \ie, $\bar{Y}_{it} = \frac{1}{t-1}\sum_{\tau=1}^{t-1}Y_{i\tau}$. 
\item {Dynamic decision rule (ARMA)}. This model considers all of the above predictors and, in addition, three autoregressive terms.
\item {LSTM. As another sequential baseline, we also implemented a recurrent neural network (\ie, a long short-term memory network, LSTM for short). This model is trained in a supervised manner analogous to the other baselines \citep[cf.][]{kraus2020deep}.}
\end{enumerate}  
These models are based on supervised learning and, in contrast to our VDC-HMMX, are thus trained via expert annotations.

\section{Model Estimation}
\label{sec:model_selection}

Here we perform a model selection to choose our preferred model specification. In particular, we compare our \model against other variants without variable-duration component as well as without copula to demonstrate the effectiveness of our proposed model. Due to space, additional details are in the Online Appendix.

\subsection{Model Variants}

We compare the following model variations that differ in the underlying personalization (all of them make use of our copula approach; additional model variations are discussed as part of the robustness checks):
\begin{quote}
\begin{enumerate}[listparindent=5cm]
\item \textbf{Variable-duration HMMX}. This corresponds to our \model but it has no copula. As such, the different health measurements are assumed as being independent. 
\item {\textbf{Copula HMMX.}} This corresponds to our \model but without a variable-duration component. This model thus follows other HMMs in management literature \citep[\eg,][]{Martino.2018,Yan.2014}.
\item {\textbf{VDC-HMMX.}} This corresponds to the proposed variant specified above.
\end{enumerate}
\end{quote}%\end{quote}

\vspace*{0.3cm}
\noindent
We base our model selection on the out-of-sample performance. For treatment planning, this yields a model that generalizes well to unseen patients. We follow the latest recommendations in Bayesian modeling \citep{BDA3} and use the out-of-sample log point-wise predictive density~(lpd). Different from the deviance information criterion~(DIC) or the Akaike information criterion~(AIC), the lpd is beneficial as it considers the whole posterior distribution. We also report approximate metrics of the lpd, namely the widely applicable information criterion~(WAIC) and the leave-one-out cross-validation information criterion~(LOO). Furthermore, we state the in-sample lpd in order to provide an indication, when overfitting occurs. All values are measured on the deviance scale such that smaller values indicate a better model fit.

%As part of our robustness checks, we also considered models with patient-individual random effects and heterogeneity in the emission component. However, the model fit was inferior. On top of that, the latter specification is further discouraged by practical considerations, as it does not control for differences in the disease dynamics but only in the reporting behavior. 

\subsection{Model Selection}

\Cref{tbl:model_selection} presents the results. The best out-of-sample lpd is obtained by the VDC-HMMX. It outperforms both the variable-duration HMMX and the copula HMMX, thereby confirming that a combination of variable-duration component and copula within an HMMX is beneficial. 

We make further observations. First, the the change from the copula HMMX to our \model (\ie, \num{862.59} in lpd) is larger than the change between a na{\"i}ve VDC-HMM without risk factors and our \model (\ie, the difference is only \num{116.00}). In other words, the improvement gained by including the variable-duration component is larger than that gained by including risk factors. This highlights the importance of relaxing the Markov property for our research and accommodate a variable-duration component. 

Second, an out-of-sample lpd of 40234.32 is obtained for the variable-duraton HMMX without copula, that is, when assuming independence among health measurements. In comparison, a value of 39966.27 is registered for our \model. This amounts to an improvement of \num{268.05}. To put this into context, this improvement is more than twice as large as the performance gained through the inclusion of risk factors (\ie, the improvement of the VDC-HMMX over a na{\"i}ve VDC-HMM amounts to only \num{116.00}). This thus highlights the importance of considering a copula approach. 

In sum, our proposed \model is to be preferred, and, hence, we draw upon the \model in all subsequent analyses.

\begin{table}[H]
	\centering
	\sisetup{parse-numbers=false}
	\TABLE{Model Selection.\label{tbl:model_selection}}
	{\scriptsize
		\begin{tabular}{l SS SS}
			\toprule
			&\multicolumn{2}{c}{Expected lpd} & \multicolumn{2}{c}{True lpd}\\
			\cmidrule(lr){2-3} 			\cmidrule(lr){4-5} 			
			Model & 
			\multicolumn{1}{c}{LOO} & 
			\multicolumn{1}{c}{WAIC} &
			\multicolumn{1}{c}{In-sample} & \multicolumn{1}{c}{Out-of-sample}\\
			\midrule
			Variable-duration HMMX (\ie, no copula) & 89742.66 & 89737.34 & 89091.24 & 40234.32 \\
			Copula HMMX (\ie, no variable-duration) & 91725.22 & 91701.80 & 90728.74 & 40828.86 \\[0.3em] 
			Proposed \model & 89402.53 & 89400.83 & 88669.65 & \bm{39966.27} \\
			\bottomrule
			\multicolumn{5}{l}{\scriptsize Lower = better (best out-of-sample performance in bold)}
		\end{tabular}	
	}{}
%\footnotesize
%\begin{tabular}{p{16cm}}
%\emph{Notes:} The table compares different model specifications with respect to personalization, that is, how patient heterogeneity is modeled. The actual model selection is based on information criteria that have been designed for Bayesian modeling \citep{BDA3}, specifically the out-of-sample lpd as a measure for how well the model generalized to unseen patients. The best performance is achieved by the VDC-HMMX. It is thus considered in all subsequent analyses.
%\end{tabular}
\end{table}

\subsection{Estimation Results}

The posterior means of the emission probabilities are listed in \Cref{tbl:emission} (\ie, marginal distributions $b_s^{(k)}$). As expected, we observe a strong lower-tail dependence between pain intensity and activity limitation. Furthermore, the three latent states match the different trajectory phases: (1)~the stable state reveals the nearly complete absence of symptoms as demonstrated by both low pain intensity and almost no activity limitation; (2)~the unstable state reflects dispersed health measurements; and (3)~the acute state shows mostly severe health measurements. Altogether, this suggests that our model has successfully encoded the trajectory phases in the latent states.

%We now characterize the latent states. This provides evidence that they reflect acute, unstable, and stable trajectory phases as defined by the trajectory framework \citep{Corbin.1988,Corbin.1991}. For this, we draw upon the selected model from above, namely the VDC-HMMX with a tail dependence copula. On this basis, we then investigate the state-observation relationship: it provides the most probable emissions in each latent state. 

\begin{table}
	\TABLE{Mean Posterior Emission Probabilities (\ie, Marginal Distributions $b_s^{(k)}$).\label{tbl:emission}}
	{\tiny
		\begin{tabular}{l cccccccc cccccccc cccccccc}
			\toprule
			&\multicolumn{24}{c}{Activity limitation} \\
			\cmidrule(lr){2-25}
			& \multicolumn{8}{c}{Stable phase} & 
			\multicolumn{8}{c}{Unstable phase} &
			\multicolumn{8}{c}{Acute phase} \\
			\cmidrule(lr){2-9} \cmidrule(lr){10-17} \cmidrule(lr){18-25}
			{Pain}&0 & 1 & 2 & 3 & 4 & 5 & 6 & 7&0 & 1 & 2 & 3 & 4 & 5 & 6 & 7	&0 & 1 & 2 & 3 & 4 & 5 & 6 & 7\\
			\midrule
			\csname @@input\endcsname mvt_emission % load file 
			\bottomrule
		\end{tabular}	
	}{\emph{Notes:} This table presents the mean posterior emission probability (in \%) for each latent state. We highlight posterior probabilities greater than \SI{1}{\percent} and \SI{5}{\percent} in light and dark gray, respectively. As a result, the latent states can be interpreted as stable, unstable, and acute trajectory phases. }
\end{table}

\section{Model Performance}
\label{sec:results}

We now utilize our proposed \model for personalizing treatment planning. 

\subsection{Accuracy of Inferred Treatment Regimens}

This section assesses the relative frequency with which clinical guidelines for choosing an effective treatment regimen are obtained. Accordingly, we compute the accuracy of how often the correct treatment regimen was identified. The data-driven approaches are evaluated based on the dynamic on-line framework. That is, they receive a health measurement at time $t$ and must then suggest a treatment regimen before the health measurement from the next week $t+1$ becomes available. As the course of symptoms is highly variable, it renders treatment planning considerably more challenging. In comparison, the ground truth annotations are provided by the medical experts who were privy to the complete disease progression for annotation and, hence, benefited from post-hoc knowledge. If one had -- hypothetically -- post-hoc knowledge of the disease progression, one could always choose the best treatment regimen in medical practice, thereby identifying the correct treatment regimen with 100\% accuracy and thus removing all costs from over-/under-treatment. However, in medical practice, this is not possible as it requires post-hoc knowledge.

We compare the performance of our \model against a range of baselines (see \Cref{sec:baselines}). \Cref{tbl:HMM_in-sample_fit} reports the F1 score (the harmonic mean of precision and recall) and the balanced accuracy (due to the uneven distribution of treatment regimens). We first study symptom-based classifiers as our prime baseline. Here the decision rules that mimic current practice register an overall balanced accuracy of \SI{72.63}{\percent} (myopic decision rule, \ie, current symptoms only) and \SI{74.25} (ARMA). These performance values already highlight challenges that health professionals face when managing chronic diseases: Identifying an effective treatment regimen without knowledge of a patient's health trajectory is difficult. Similar findings from qualitative research have been reported earlier \citep[\eg,][]{Burton.2000,Corbin.1991}. 

In comparison, our data-driven model attains an overall balanced accuracy of \SI{83.65}{\percent}. Hence, it bolsters the overall accuracy by \num{9.49} percentage points. For all baseline decision rules, the improvement is statistically significant at the 0.001 significance threshold. In particular, our data-driven model is more accurate when acute treatment regimens are needed. Hence, it facilitates care provision for patients with severe conditions. 

The LSTM registers a balanced accuracy of 81.33 and an F1 score of 73.91. Hence, it outperforms the other baselines mimicking current practice but, in comparison to our \model, it is still inferior by a large margin. On top of that, the long short-term memory is regarded as a black-box and, based on discussions with healthcare professionals, its use was discouraged. In our case, the tuned network architecture has \num{6703} parameters, which is \num{43} times larger than the parsimonious formulation of the \model. Due to this fact, the use of such a recurrent neural network is further limited in practice.

\begin{table}[H]
\centering
	\TABLE{Accuracy in Inferring Treatment Regimens.\label{tbl:HMM_in-sample_fit}}
	{\scriptsize
		\begin{tabular}{l r SSSS SSSS}
			\toprule
			& &\multicolumn{4}{c}{F1 score} & \multicolumn{4}{c}{Balanced accuracy} \\
			\cmidrule(lr){3-6}\cmidrule(lr){7-10}
			Model &  & {Overall} & {Stable}  & {Unstable} & {Acute} & {Overall} & {Stable}  & {Unstable} & {Acute} \\
			\midrule
      Baselines\\
			\qquad Myopic decision rule (symptoms only)  & &65.86 &   84.20 &  43.32 &  70.06 &  72.63 &  76.25 &  62.82 &  78.81 \\
			\qquad Myopic decision rule (risk factors) & & 67.47 &   84.96 &   45.62 &   71.84  &  73.66  &  77.53  &  63.86&    79.58\\			
			\qquad Dynamic decision rule (moving average) & & 67.23 &   85.54  &  52.82  &  63.33 &   74.00  &  79.02 &   67.59   & 75.40 \\
			\qquad Dynamic decision rule (ARMA) & & 67.24 &   85.81  &  53.48 &   62.43   & 74.25  &  79.54  &  67.95  &  75.25  \\[0.3em]
%			Neural networks\\
			\qquad LSTM & & 73.91 &   91.00 &  64.30 &  66.43 &  81.33  &  88.81 &  74.68 &  80.40 \\[0.3em]
			Proposed \model & & 77.02 &   90.69 &  66.14  & 74.24  & 83.65 &  88.95 &  76.02 &  85.98\\
			\midrule
			Improvement over best baseline (in percentage points)  & & 9.78 &   4.88 &  12.66  & 3.68  & 9.40 &  9.41 &  8.07 &  6.99 \\
			\bottomrule
			\multicolumn{10}{l}{Stated: macro-averaged overall and class-specific performance; balanced accuracy in \%.}
		\end{tabular}	
	}{
	\footnotesize
}
\end{table}

\subsection{Cost-Effectiveness}

\Cref{tbl:costs} compares the effectiveness based on the improvement in QALY over \textquote{no treatment} across different models. As the results show, our model outperforms the best baseline by 12.97\,percentage points. This directly reflects benefits for patients' health provided by our data-driven approach to treatment planning. We further provide a cost analysis of over- and under-treatment. Hence, for the given decision rules, we accumulate over the absolute deviation in costs between suggested and expert-annotated treatment regimens, thereby following common guidelines in chronic care \citep{Larsen.2017}. Compared to our model, the LSTM is fairly similar in terms of total costs but has a substantially lower QALY improvement. The different interpretable baseline models yield fairly similar annual over-/under-treatment costs of, on average, USD~37.69 per patient. For comparison, our model implies annual over-/under-treatment costs of USD~\num{23.80} per patient. This amounts to an overall improvement of \SI{35.71}{\percent}. In particular, our approach saves costs stemming from under-treatment, whereby inadequate treatment regimens are selected that are not suited for managing the corresponding trajectory phase. 
 Note that better treatment regimens implicitly entail additional cost savings, such as from reduced side-effects.

\begin{table}[H]
	\centering
	\TABLE{Model Comparison by Cost-Effectiveness .\label{tbl:costs}}	
	{\scriptsize
		\begin{tabular}{l SSS S}
			\toprule
			& \multicolumn{3}{c}{Cost} & \multicolumn{1}{c}{Effectiveness} \\
			\cmidrule(lr){2-4} \cmidrule(lr){5-5}
			Model & {Total} & {Under-treatment}  & {Over-treatment} & {QALY (improvement)} \\
			\midrule
			Baselines\\
			\qquad Myopic decision rule (symptoms only) & 39.48 & 36.76 &  2.73 & \SI{72.56}{\percent}  \\
			\qquad Myopic decision rule~(risk factors) & 37.26 & 34.97 &  2.29 & \SI{73.26}{\percent} \\
			\qquad Dynamic decision rule (moving average) & 37.38 & 32.98 & 4.39 & \SI{75.61}{\percent} \\			
			\qquad Dynamic decision rule (ARMA) & 37.02 & 32.15 & 4.89 & \SI{76.32}{\percent} \\[0.3em]
%			Neural networks\\
			\qquad LSTM & 25.39 & 15.51 & 9.88 & \SI{76.31}{\percent}  \\[0.3em]
			%\midrule
			Proposed VDC-HMMX & 23.80 & 12.53 & 11.26 & \SI{89.29}{\percent}\\
%			\midrule
%			Improvement over best baseline (in \%) & 35.71 & 61.02 & -130.32 & {12.97\,p.p.} \\
			\bottomrule
			\multicolumn{5}{l}{Stated: absolute deviation between costs from suggested and needed treatment regimen}\\
			\multicolumn{5}{l}{per patient and annum  (in USD); improvement in QALY over a \textquote{no treatment} strategy (in\,\%)}
		\end{tabular}	
	}{
		\footnotesize
	}
	\vspace{-1cm}
\end{table}

\section{Discussion}
\label{sec:discussion}

\subsection{Implications for Health Management}

In order to provide effective care, chronic disease management aims at managing the trajectory of chronic diseases. Each trajectory phase is characterized by different disease dynamics and thus requires a tailored treatment regimen. This has been formalized in the trajectory framework  \citep{Corbin.1991} and is recommended to health professionals as best practice in chronic disease management \citep{Larsen.2017}. Yet applying the trajectory framework in clinical practice is challenging, as symptoms and the underlying trajectory are only stochastically related. This renders the correct identification of a patient's trajectory phase, and thus of effective treatment regimens, challenging \citep{Burton.2000}. As a remedy, this paper develops a data-driven approach, suggesting treatment regimens that are personalized to a patient's individual health trajectory.

Our work contributes to chronic disease management by demonstrating the operational value of longitudinal monitoring \citep[cf.][]{JMIR}. It allows health professionals to manage the disease progression throughout all trajectory phases. This is especially helpful for chronic diseases as these are characterized by a recurrent and long-lasting progression, which requires regular adjustments to treatment plans. Without longitudinal knowledge, treatment decisions can only be based on current symptoms rather than the disease dynamics behind them (\eg, an aggressive, persistent, or relapsing course). In the role of providing decision support to health professionals, our \model model appears highly effective as it recovers the otherwise unobservable trajectory phases. As in other management applications, it also suggests intervention points when the treatment regimens need adjustments.  

Our model was specifically designed to meet the requirements of professionals and researchers in the healthcare sector who demand a high degree of interpretability. Instead of using black-box models such as deep neural networks, we accomplished this via a parsimonious formulation. Specifically, we followed the trajectory framework whereby the complex course of multi-dimensional health measurements is mapped onto three states with clinically relevant interpretations. For this purpose, we propose a novel {variable-duration copula HMM} called \model.

\subsection{Limitations and Potential for Future Research}

We are aware that our work is subject to limitations that hold for the majority of studies in healthcare research. Our study involved a large, heterogeneous cohort of patients with standardized examinations. Nevertheless, other symptoms and risk factors might be of interest for future research. Furthermore, our approach follows best practice in chronic care, where the focus is not on a cure but on stabilizing the trajectory \citep{Larsen.2017}. The objective of our work was to aid health professionals in managing chronic diseases throughout the complete trajectory. Hence, we make recommendations as to which treatment regimen (\ie, acute, unstable, or stable) should be selected and when it should be updated. Once a treatment regimen has been selected, its design might be subject to further personalization. This latter dimension, customizing a given treatment regimen such as returned from our model, has already been addressed in earlier studies (see \Cref{sec:background}).

Our data-driven approach to personalized treatment planning is not limited to low back pain, but is applicable to other chronic diseases that follow the trajectory framework \citep{Corbin.1988,Corbin.1991}. To do so, one simply has to update the health measurements in our model to disease-specific symptoms (\eg, one might need to consider a different time lag for treatment variables). For instance, multiple sclerosis could be monitored by the severity of blurred vision; neurodermatitis by the size of blisters; rheumatoid arthritis by pain; migraine, epilepsy, and bipolar disorder by the intensity and frequency of attacks.

\section{Conclusion}
\label{sec:conclusion}

Chronic diseases exhibit a recurrent trajectory with varying needs for care. Here, we address the important decision-making problem of managing treatment regimens through a patient's complete trajectory, where dynamic updates to the treatment regimens are needed. Specifically, we develop a novel variable-duration copula HMM called \model. This allows health professionals to manage disease dynamics (rather than symptoms). Thereby, we aim to provide effective care and reduce over- and under-treatment. 

\vspace{-0.3cm}
%\ACKNOWLEDGMENT{
\begin{table}[H]
\SingleSpacedXI\footnotesize
\begin{tabular}{p{\linewidth}}
\textbf{Acknowledgments.} SF acknowledges funding from the Swiss National Science Foundation (SNSF) as part of grant 186932.
\end{tabular}
\end{table}%}
\vspace{-0.4cm}

% References here (outcomment the appropriate case) 

% CASE 1: BiBTeX used to constantly update the references 
%   (while the paper is being written).
%\OneAndAHalfSpacedXI
\SingleSpacedXI
\footnotesize
\bibliographystyle{informs2014} % outcomment this and next line in Case 1
\bibliography{literature} % if more than one, comma separated

% CASE 2: BiBTeX used to generate mypaper.bbl (to be further fine tuned)
%\input{mypaper.bbl} % outcomment this line in Case 2

\newpage
\begin{APPENDICES}
\begin{center}
\LARGE Online Supplements
\end{center}

\section{Model Estimation}
\label{appendix:model_estimation}

% overview

The model parameters are estimated using a Bayesian approach as follows. We use Markov chain Monte Carlo (MCMC) in order to directly sample from the posterior distribution of the model parameters. For this reason, we first derived the likelihood $L$ for the complete HMM with variable-duration component and copula approach. Our implementation draws upon recent advances in Bayesian estimations \citep{BDA3}, namely, the Hamiltonian Monte Carlo sampler together with the No-U-Turn (NUTS) technique from Stan. This approach differs from other estimation techniques, specifically the Metropolis-Hastings algorithm or maximum likelihood estimation. In contrast to these methods, our estimation approach leverages an explicit derivation of likelihood in order to direct sample from the posterior distribution. This is known to be considerably more efficient and, together with Hamiltonian Monte Carlo, requires fewer chains/iterations by a several orders of magnitude \citep{BDA3}. As a result, we do not need to derive a maximum likelihood approach, an expectation-maximization algorithm, or a Metropolis-Hastings scheme; instead, the derived likelihood $L$ is sufficient. 

We now detail the procedure for estimating the hidden Markov models based on their log-likelihood. For this purpose, we first need to specify appropriate priors for all model parameters in order to then sample from the posterior distribution. We choose weakly informative priors for all model parameters as follows: 
\begin{enumerate}
\item \emph{\textbf{Initial state distribution.}} The initial state distribution $\pi$ of the trajectories is given by $\pi_s = P(S_{i1} = s)$ for $s = 1, \ldots, \varsigma$. Here we assume $\pi\sim\textup{Dir}(1,\ldots,1)$, which is equivalent to a uniform distribution over the open standard $(\varsigma - 1)$-simplex.	
\item \emph{\textbf{Transitions.}} For the na{\"i}ve (VDC-)HMM, we simply choose a Dirichlet prior with all parameters equal to $1$, except that we set the diagonal elements to $\varsigma$ in order to penalize frequent switching between the latent states. The covariates from the (VDC-)HMMX models are first centered and then subject to a QR decomposition, yielding a re-parametrization $\tilde{\beta}_{jk}=R^{-1}\beta_{jk}$ for reasons of numerical stability. We then place a standard normal prior on $\tilde{\beta}_{jk}$ and a zero-mean normal distribution with standard deviation $10$ on the intercepts $\delta_{jk}$ for $j,k \in \{ 1, \ldots, \varsigma \}$.
For the parameter $\omega_{jl}$, we chose standard normal priors. 
\item \emph{\textbf{Emissions.}} The choice of the emission prior depends on the type of the observed health variables. Continuous variables (\eg, pulse or blood sugar levels) could be modeled as normal distributions, while Likert-based values, as in our case, are commonly assumed to follow a truncated Poisson distribution. For instance, pain levels have been found to follow such a truncated Poisson distribution \citep{Goulet.2017}. Here the parameters specify the distribution means $\lambda_s^{(1)}$, $\lambda_s^{(2)}$, which are given a zero mean normal distribution with standard deviation of $5$ truncated at zero as prior. Furthermore, it is beneficial to impose the constraint $\lambda_1^{(1)}<\ldots<\lambda_{\varsigma}^{(1)}$ which orders the states by severity. This leaves the findings unchanged, but avoids label-switching and ensures model identification across individual runs.  	
\item \emph{\textbf{Copula.}} The copula parameter $\nu$ inside $C_s$ is required to be greater than or equal to one. Thus, they are rewritten as $\nu = 1 + \tilde{\nu}$ such that a standard normal prior truncated at zero is placed on $\tilde{\nu}$ for each state $s$. 
\end{enumerate} 

% ordering

As stated above, we enforce an ordering on the parameters corresponding to the first margin of the observations, \ie, $\lambda_1^{(1)} <\ldots <\lambda_{\varsigma}^{(1)}$. This serves two purposes: First, this ordering ensures that the first trajectory phase~(\ie, stable) is always identified with the lowest values in the first margin. Second, the ordering addresses the problem of label-switching common in Bayesian mixture models \citep{Jasra.2005} and has been suggested in prior research as a remedy \citep{Yan.2018}.

For each model, we ran two Markov chains, each with a total of \num[group-minimum-digits=3]{3500} iterations, of which we discarded the initial \num[group-minimum-digits=3]{1000} iterations as part of a warm-up, yielding a total of \num[group-minimum-digits=3]{5000} samples for each model. We subsequently ensured the convergence of chains by inspecting the individual trace plots and the number of effective samples for each parameter. We further checked that the Gelman-Rubin convergence index for each parameter was below \num{1.02}. We also validated our model design by testing whether we can retrieve the parameters from simulated data. All checks yielded the desired outcomes.

\section{Model Variants}
\label{appendix:model_overview}

We compare the following model variations that differ in the underlying personalization (all of them make use of our copula approach; additional model variations are discussed as part of the robustness checks):
\begin{quote}
\begin{enumerate}[listparindent=5cm]
\item {\textbf{Na{\"i}ve HMM.}} The na{\"i}ve HMM lacks both the variable-duration component, as well as $x_{it}$ with other sources of heterogeneity. As a result, its transition probability does not depend on the latent state duration $d_{it}(s)$ and is thus stationary. Mathematically, this presents a special case of the VDC-HMM for which $\omega_{jl}\equiv0$ for $j,l\in\lbrace1,\ldots,\varsigma\rbrace$.	
\item {\textbf{HMMX.}} This is like the na{\"i}ve HMM without a variable-duration component, but the covariates $x_{it}$ (\ie, treatments and various risk factors such as gender) are included. This model thus follows other HMMs in management literature \citep[\eg,][]{Martino.2018,Yan.2014}.
\item {\textbf{Na{\"i}ve VDC-HMM.}} The na{\"i}ve VDC-HMM overcomes limitations of other HMMs \citep[\eg,][]{Netzer.2008,%Singh.2011,
Yan.2014}, so that the transition probability is affected by the duration $d_{it}(s)$ spent in a latent state. Note that the duration $d_{it}(s)$ is itself latent, and, hence, $d_{it}(s)$ cannot simply be inserted into the model. The na{\"i}ve VDC-HMM \citep[][p.\,622]{Murphy.2012} provides a baseline in which other sources of heterogeneity, such as patient-specific risk factors, are ignored.   
\item {\textbf{VDC-HMMX.}} This variant was specified above. It models the transition probabilities based on the state duration and includes covariates $x_{it}$ (\ie, treatment data and patient-specific risk factors). This allows the propensity of moving between states to differ based on a patient's risk profile.
\item \textbf{Patient-individual VDC-HMM.} We also consider two additional models with patient-specific random effects. In the first variant, the first-order VDC-HMMX is extended by placing a hierarchical prior $\mathcal{N}(\mu_{jl},\sigma_{jl}^2)$ on the intercepts $\delta_{jl}$ for $j,l\in\lbrace1,\ldots,\varsigma\rbrace$, $j\neq l$. Second, we use patient-specific random effects for $\omega_{jl}$. That is, we place a hierarchical prior on each parameter, \ie, $\omega_{jl}\sim\mathcal{N}(\mu_{jl}',{\sigma_{jl}'}^2)$. At test time, the parameters from the random effects are not known and we thus set them to the respective hierarchical mean. 
\end{enumerate}
\end{quote}%\end{quote}

\vspace*{0.3cm}
\noindent
As part of our robustness checks, we also considered models with patient-individual random effects and heterogeneity in the emission component. However, the model fit was inferior. On top of that, the latter specification is further discouraged by practical considerations, as it does not control for differences in the disease dynamics but only in the reporting behavior.

\section{Model Selection}
\label{appendix:model_selection}

We first choose (1)~the number of latent states, (2)~the model specification (which controls for patient heterogeneity), and (3)~the copula (which handles interdependent health measurements). The selected model is then interpreted to show that the latent states match the characteristics of the trajectory phases from the trajectory framework. The model variants for comparison are listed in Appendix~\ref{appendix:model_overview}.

We base our model selection on the out-of-sample performance. For treatment planning, this yields a model that generalizes well to unseen patients. We follow the latest recommendations in Bayesian modeling \citep{BDA3} and use the out-of-sample log point-wise predictive density~(lpd). Different from the deviance information criterion~(DIC) or the Akaike information criterion~(AIC), the lpd is beneficial as it considers the whole posterior distribution. We also report approximate metrics of the lpd, namely the widely applicable information criterion~(WAIC) and the leave-one-out cross-validation information criterion~(LOO). Furthermore, we state the in-sample lpd in order to provide an indication, when overfitting occurs. All values are measured on the deviance scale such that smaller values indicate a better model fit.

\subsection{Determining the Number of Latent States}

We first determine the number of latent states, $\varsigma$, that is best in describing the course of the disease. Here we re-estimate VDC-HMMXs with different $\varsigma$. \Cref{tbl:number_latent_states} finds the best out-of-sample lpd for three latent states.  $\varsigma = 3$ latent states were also conceptualized in the trajectory framework \citep{Corbin.1991}, thereby yielding empirical confirmation. Hence, all subsequent analyses are based on three latent states. 

\begin{table}[H]
	\centering
	\sisetup{parse-numbers=false}
	\TABLE{Number of Latent States.\label{tbl:number_latent_states}}
	{\footnotesize
		\begin{tabular}{l SS SS}
			\toprule
			&\multicolumn{2}{c}{Expected lpd}  & \multicolumn{2}{c}{True lpd}\\
			\cmidrule(lr){2-3} 			 \cmidrule(lr){4-5}			
			\#States $\varsigma$ & 
			\multicolumn{1}{c}{LOO} & 
			\multicolumn{1}{c}{WAIC} & 
			\multicolumn{1}{c}{In-sample} & \multicolumn{1}{c}{Out-of-sample}\\
			\midrule
			1 & 168962.29 & 168958.27 & 168608.95 & 73204.05 \\  
			2 & 103952.57 & 103951.31& 103711.85 & 46126.47 \\ 
			3 & 89402.53 & 89400.83 & 88669.65 & \bm{39966.27} \\ 
      4 & 92576.07 & 92572.56  & 92270.17 & 41191.07 \\ 
			\bottomrule	
		\end{tabular}
	}
	{}
	\footnotesize
	\begin{tabular}{p{16cm}} \emph{Notes:} This table determines the optimal number of latent states $\varsigma$. For this purpose, the VDC-HMMX is estimated with a different number of latent states. The actual model selection is based on information criteria that have been designed for Bayesian modeling \citep{BDA3}, specifically the out-of-sample lpd (and its approximations). By considering the out-of-sample performance, we quantify the ability of the model to generalize to unseen patients. The results confirm the trajectory framework \citep{Corbin.1991}, whereby $\varsigma = 3$ should be preferred. 
\end{tabular}
\end{table}

\subsection{Model Choice for Specifying Patient Heterogeneity}

We now consider how to personalize the hidden Markov model to the individual risk profiles of patients. For this purpose, different models are estimated that vary in their model specification, particularly with respect to how they control for the heterogeneity in patients' risk profiles. All models draw upon the copula approach. Subsequently, the model with the best fit is selected.

\begin{table}[H]
	\centering
	\sisetup{parse-numbers=false}
	\TABLE{Model Selection for Handling Patient Heterogeneity.\label{tbl:VDHMM_model_choice}}
	{\footnotesize
		\begin{tabular}{l SS SS}
			\toprule
			&\multicolumn{2}{c}{Expected lpd} & \multicolumn{2}{c}{True lpd}\\
			\cmidrule(lr){2-3} 			\cmidrule(lr){4-5} 			
			Model & 
			\multicolumn{1}{c}{LOO} & 
			\multicolumn{1}{c}{WAIC} &
			\multicolumn{1}{c}{In-sample} & \multicolumn{1}{c}{Out-of-sample}\\
			\midrule
			\csname @@input\endcsname vdhmm_model_choice % load file			
			\bottomrule
		\end{tabular}	
	}{}
\footnotesize
\begin{tabular}{p{16cm}}
\emph{Notes:} The table compares different model specifications with respect to personalization, that is, how patient heterogeneity is modeled. The actual model selection is based on information criteria that have been designed for Bayesian modeling \citep{BDA3}, specifically the out-of-sample lpd as a measure for how well the model generalized to unseen patients. The best performance is achieved by the VDC-HMMX. It is thus considered in all subsequent analyses.
\end{tabular}
\end{table}

\Cref{tbl:VDHMM_model_choice} presents the results. The best out-of-sample lpd is obtained by the VDC-HMMX. In comparison to the na{\"i}ve VDC-HMM, the model fit improves when considering the between-patient heterogeneity in $x_{\it}$ (\ie, treatment and patient risk factors). This highlights the need for personalization by modeling the between-patient heterogeneity. 

Based on \Cref{tbl:VDHMM_model_choice}, we make further observations. First, random effects, as in the patient-individual models, result in overfitting. While the patient-individual models improve in terms of in-sample fit, their out-of-sample performance is sub-par. In fact, both patient-individual models attain an out-of-sample lpd that is even below that of the na{\"i}ve VDC-HMM. Second, the na{\"i}ve HMM yields the poorest performance among all considered models. This holds true across all metrics. Third, the change from the HMMX to the VDC-HMMX (\ie, \num{862.59} in lpd) is larger than the change between the na{\"i}ve VDC-HMM and the VDC-HMMX (\ie, the difference is only \num{116.00}). In other words, the improvement gained by including the variable-duration component is larger than that gained by including risk factors. This highlights the importance of relaxing the Markov property for our research. Hence, we draw upon the VDC-HMMX in all subsequent analyses.

\subsection{Copula Selection for Modeling Dependence Structure}

Symptoms are expected to co-occur and, hence, we evaluate approaches for modeling the dependence structure among health measurements. \Cref{tbl:VDHMM_copula_choice} reports the results. An out-of-sample lpd of 40234.32 is obtained when assuming independence, that is, when we specify our model without copula. In comparison, a value of only 39966.27 is registered by the survival Gumbel copula. This amounts to an improvement of \num{268.05}. To put this into context, this improvement is more than twice as large as the performance gained through the inclusion of risk factors (\ie, the improvement of the VDC-HMMX over the na{\"i}ve VDC-HMM amounts to only \num{116.00}). We further note that the survival Gumbel copula is superior not only in terms of the out-of-sample lpd, but also for the other information criteria. In sum, the results demonstrate that accounting for a tail dependence via the survival Gumbel copula is highly effective.

\begin{table}[H]
	\centering
	\sisetup{parse-numbers=false}
	\TABLE{Copula Selection for Handling Dependence Structure among Health Measurements.\label{tbl:VDHMM_copula_choice}}
	{\footnotesize
		\begin{tabular}{l SS SS}
			\toprule
			&\multicolumn{2}{c}{Expected lpd} & \multicolumn{2}{c}{True lpd}\\
			\cmidrule(lr){2-3} 			\cmidrule(lr){4-5} 			
			Copula & 
			\multicolumn{1}{c}{LOO} & 
			\multicolumn{1}{c}{WAIC} &
			\multicolumn{1}{c}{In-sample} & \multicolumn{1}{c}{Out-of-sample}\\
			\midrule
			Independence (\ie, no copula) & 89742.66 & 89737.34 & 89091.24 & 40234.32 \\
			Independence (with correlated prior) & 89815.62 & 89813.24 & 89154.44 & 40250.78\\[0.3em]
			Symmetric dependence (Ali-Mikhail-Haq copula) & 89439.30 & 89434.96 & 88720.96 & 39999.35\\
			Tail dependence (Clayton copula) & 89469.89 & 89482.82 & 88710.3 &  40018.36 \\
			Tail dependence (survival Gumbel copula) & 89402.53 & 89400.83 & 88669.65 & \bm{39966.27} \\
			\bottomrule
		\end{tabular}	
	}{}
	\footnotesize
	\begin{tabular}{p{16cm}}
		\emph{Notes:} The table compares different copulas across different information criteria for Bayesian modeling. In order to quantify the performance when the model is applied to unseen patients, our selection is based on the out-of-sample lpd.  
	\end{tabular}
\end{table}

\Cref{tbl:VDHMM_copula_choice} lists further baselines for comparison. We estimate another model without copula but where the priors are correlated. Specifically, for the parameters $\lambda_s^{(1)},\lambda_s^{(2)}$, we assume a multivariate normal distribution with covariance matrix $\Sigma_s$ and zero-means as prior. This result is largely on par with a complete independence~(which is included as a special case where $\Sigma_s$ is diagonal) approach from above and, therefore, inferior to our dependence structure. The Ali-Mikhail-Haq copula with a symmetric dependence improves upon a model without copula but underperforms in comparison to our model with tail dependence. Therefore, all subsequent analyses are based on the survival Gumbel copula in order to accommodate a tail dependence among health measurements. While the Clayton copula also models tail dependence, it underperforms the survival Gumbel copula, suggesting that it captures the observed dependence to a lesser extent.

\section{Robustness Checks}
\label{appendix:robustness_checks}

An extensive series of robustness checks was performed in order to validate our results. We summarize the main findings in the following. (1)~We repeated the above analysis with univariate emissions. As expected, the results are inferior to multivariate emissions. The same held true when we applied dimensionality reduction via a principal component analysis. (2)~We encoded different structures in the transition matrix (\ie, funnel structure, absorbing state), but, consistent with the trajectory framework, these resulted in an inferior model fit. (3)~We included risk factors in the emission, though with an inferior model fit. This was expected, as such a model no longer controls for heterogeneity in the disease dynamics but rather in the reporting behavior. (4)~Patient-individual random effects in the emissions led to overfitting. (5)~We used the empirical distribution in the multivariate emission, but the prediction was subject to overfitting. (6)~We utilized the VineCopula package for R in order to compare different copulas numerically. Here we experimented with the following set of copulas: tawn type II, BB7, Frank, Clayton, and Joe. The numerical findings supported the choice of the survival Gumbel copula. (7)~Removing treatments or the current health measurements from the risk factors $x_{it}$ resulted in an inferior fit. (8)~We used machine learning to map $S_{it}^\ast$ onto $\phi_{it}$ with additional risk variables, but this led to overfitting.

\end{APPENDICES}

\end{document}